\documentstyle[epsf]{mn}
\input epsf.sty
\def\hmpc{~h$^{-1}$ Mpc~}
\def\htre{~h$^{-3}$ Mpc$^3$~}

\title[A study of the core of the Shapley Concentration IV]
 { A study of the core of the Shapley Concentration: \\
 IV. Distribution of intercluster galaxies and supercluster properties
\thanks{based on observations collected at the European Southern
Observatory, La Silla, Chile.} }
\author[S. Bardelli et al.]{
S. Bardelli$^{1,2}$,
E. Zucca$^{1}$,
G. Zamorani$^{1,3}$,
L. Moscardini$^{4}$
\&
R. Scaramella$^{5}$
\\ $^1$ Osservatorio Astronomico di Bologna, 
via Ranzani 1, I--40127 Bologna, Italy
\\ $^2$ Osservatorio Astronomico di Trieste, 
via Tiepolo 11, I--34131 Trieste, Italy
\\ $^3$ Istituto di Radioastronomia del CNR, 
via Gobetti 101, I--40129 Bologna, Italy
\\$^4$ Dipartimento di Astronomia, Universit\`a di Padova, 
vicolo dell'Osservatorio 5, I-35122 Padova, Italy
\\$^5$ Osservatorio Astronomico di Roma,
via Osservatorio 2, I--00040 Monteporzio Catone (RM), Italy 
\\ E-mail: bardelli@astbo3.bo.astro.it
}
\date{Received 00 - 00 - 0000; accepted 00 - 00 - 0000}
\begin{document}
\maketitle
\begin{abstract}

We present the results of a redshift survey of intercluster galaxies in the 
central region of the Shapley Concentration supercluster, aimed at determining 
the distribution of galaxies in between obvious overdensities. Our sample is
formed by 442 new redshifts, mainly in the $b_J$ magnitude range $17 - 18.8$.
Adding the data from our redshift surveys on the A3558 and A3528 complexes,
which are close to the geometrical centre of this supercluster, we obtain a
total sample of $\sim 2000$ radial velocities. 
\\
The average velocity of the observed intercluster galaxies in the Shapley
Concentration appears to be a function of their ($\alpha$, $\delta$) position,  
and it can be fitted by a plane in the three--dimensional space ($\alpha$, 
$\delta$, $v$): the distribution of the galaxy distances around the best fit 
plane is described by a Gaussian with dispersion $3.8$ \hmpc. 
\\ 
Using the 1440 galaxies of our total sample in the magnitude range 
$17 - 18.8$, we reconstructed the density profile 
in the central part of the Shapley Concetration; moreover we detected another 
significant overdensity at $\sim 30000$ km/s (dubbed S300). 
\\
We estimate the total overdensity in galaxies, the mass and the dynamical
state of these structures, and discuss the effect of considering a bias 
between the galaxy distribution and the underlying matter. 
\\ 
The estimated total overdensity in galaxies of these two structures 
is $( N / \bar{N}) \sim 11.3$ on scale of $10.1$ \hmpc
for the Shapley Concentration and $( N / \bar{N}) \sim 2.9$
on scale of $24.8$ \hmpc for S300. If light traces the mass distribution, 
the corresponding masses are 
$1.4\times 10^{16}$ $\Omega_o$ h$^{-1}$ M$_{\odot}$ and 
$5.1\times 10^{16}$ $\Omega_o$ h$^{-1}$ M$_{\odot}$ for Shapley Concentration
and S300, respectively.
\\
A dynamical analysis suggests that, if light traces 
mass and $\Omega_o=1$, the Shapley Concentration already reached its turnaround 
radius and has started to collapse: the final collapse will happen in $\sim 3
\ 10^9$ h$^{-1}$ yrs. 
\\
We find an indication that the value of the bias between clusters and galaxies
in the Shapley Concentration is higher that that reported in literature,
confirming the impression that this supercluster is very rich in clusters.
\\ 
Finally, from the comparison with some theoretical scenarios, we find that
the existence of the Shapley Concentration is more consistent with the 
predictions of the models with a matter density parameter $<1$, such as open 
CDM and $\Lambda$CDM. 
\end{abstract}

%
%
\section{Introduction}

Superclusters of galaxies are the largest coherent and massive structures 
known in the Universe. These objects are crucial in cosmology because their
extreme characteristics set topological and physical constraints on the 
models for galaxy and cluster formation. For this reason, there have been great 
efforts in order to estimate the extension, shape, mass and dynamical state
of these entities. 

The most direct way to infer these quantities is to perform a 
redshift survey in order to map the galaxy distribution in the structure and
in its surrounding field, but this method requires a large amount of telescope 
time and limits the number of targets. Up to now only few of such objects have
been studied in detail: the Great Wall (Geller \& Huchra 1989; Ramella et al. 
1992; Dell'Antonio et al. 1996), the Perseus--Pisces (Haynes \& Giovanelli 
1986), the Hercules (Barmby \& Huchra 1998), and the Corona Borealis 
(Postman, Geller \& Huchra 1988; Small et al. 1998a) superclusters. 

Hudson (1993a) computed the mean overdensity of the superclusters found in his
redshift compilation ($cz < 8000$ km/s) and estimated that the mean galaxy 
density excess for these structures is $\sim 3-5$, on scales of the order of
$30-50$ \hmpc (see also Chincarini et al. 1992). 

Another way to study superclusters is to detect structures delineated by the 
distribution of clusters. This method does not require large redshift surveys 
and relies on the assumption that clusters and galaxies trace in the same
way the underlying matter distribution.
The Shapley Concentration (Scaramella et al. 1989) 
stands out as the richest system of Abell clusters
of the list of Zucca et al. (1993), at every density excess. In particular, 
at a density contrast of $\sim 2$, it contains 25 members (at mean
velocity of $\sim 14000$ km/s) contained in a box of comoving size 
($\alpha \times \delta \times D$) $32\times 55\times 100$ \hmpc (hereafter
$h=H_o/100$). 
As a comparison, at the same density contrast the Great Attractor, which is 
the largest mass condensation within $80$ \hmpc (Lynden--Bell et al. 1988),
has only 6 members, while Corona Borealis and Hercules are formed by 10
and 8 clusters, respectively.

Scaramella et al. (1989) suggested that this supercluster could be responsible 
of a significant fraction of the acceleration acting on the Local Group, by
adding 
its dynamical pull to that of the Great Attractor, which lies approximately in 
the same direction on the sky at a distance of $\sim 4000$ km/s.  
Further studies based on the dipole of the distribution of the Abell clusters 
(Scaramella et al. 1991, Plionis \& Valdarnini 1991) confirmed the suggestion
that large scales are important for cosmic flows.  

Attempts to determine the density excess and the mass of the Shapley 
Concentration have been made by Raychaudhury et al. (1991), Scaramella et al.
(1994), Quintana et al. (1995) and Ettori et al. (1997): these authors 
obtained mass estimates of $\sim 10^{16}$ h$^{-1}$ M$_{\odot}$. 
These works are essentially based on 
estimates of the cluster masses, neglecting the contribution of the 
intercluster matter. Quintana et al. (1995) gave also a value for the
total mass of the supercluster of $\sim 5 \times 10^{16}$ h$^{-1}$ M$_{\odot}$ 
on a scale of $\sim 17$ \hmpc, using the virial mass estimator applied to the 
distribution of clusters. They computed the
velocity dispersion and the virial radius using the mean velocity and the
bi-dimensional positions of clusters:
this result could be biased if the physical elongation along the line of sight
of the supercluster is not negligible. 

Studying the cluster distribution, at high density contrast the Shapley 
Concentration is 
characterized by three dense complexes of interacting clusters, dominated by 
A3558, A3528 and A3571, respectively. 
At lower density contrast these three systems connect to each other
through a large cloud of clusters.
The clusters in the A3558 and A3528 complexes appear to be aligned 
perpendicularly to the line of sight and approximately at the same distance, 
thus suggesting the presence of an elongated underlying structure. 
However, when the whole supercluster is considered, it appears 
to be extended along the line of sight (see e.g. figure 4 in Tully
et al. 1992 and Zucca et al. 1993).

The distribution of clusters inside this supercluster is quite well studied;
on the contrary, little is known about the distribution of the intercluster
galaxies. Very recently, Drinkwater et al. (1999) performed a redshift survey
limited to the magnitude $R<16$ in the southern central part of the Shapley 
Concentration, finding evidences of sheets of galaxies connecting the clusters.

The study of the properties of intercluster galaxies is very important in order 
to assess the physical reality and extension of the structure and to determine 
if galaxies and clusters trace the matter distribution in the same way. 
In this context, we are carrying on a long term multiwavelength study of the
central part of the Shapley Concentration, studying both cluster and
intercluster galaxies. In this paper we present the results of an intercluster
galaxy redshift survey, from which we obtained $\sim 450$ new
velocity determinations, and the analysis of the whole supercluster properties.

The plan of the paper is the following: in Sect. 2 we describe the results of 
the intercluster galaxy survey and in
Sect. 3 we present the characteristics of the cluster galaxy samples. 
In Sect. 4 we analyze the galaxy distribution and in Sect. 5 we describe the 
methods adopted to recover the density profile and the mass of the whole 
structure. In Sect. 6 we derive the overdensities associated to intercluster 
and cluster galaxies, we estimate the mass of the supercluster and its 
dynamical state and we discuss our results. 
Finally in Sect. 7 we provide the summary.
In the following, the values $H_o = 100$ km/s Mpc$^{-1}$ and $q_o=0.5$ will
be adopted.   

\begin{figure*}
\centering
\leavevmode
\epsfysize=8.5cm
\epsfxsize=0.5\hsize \epsfbox{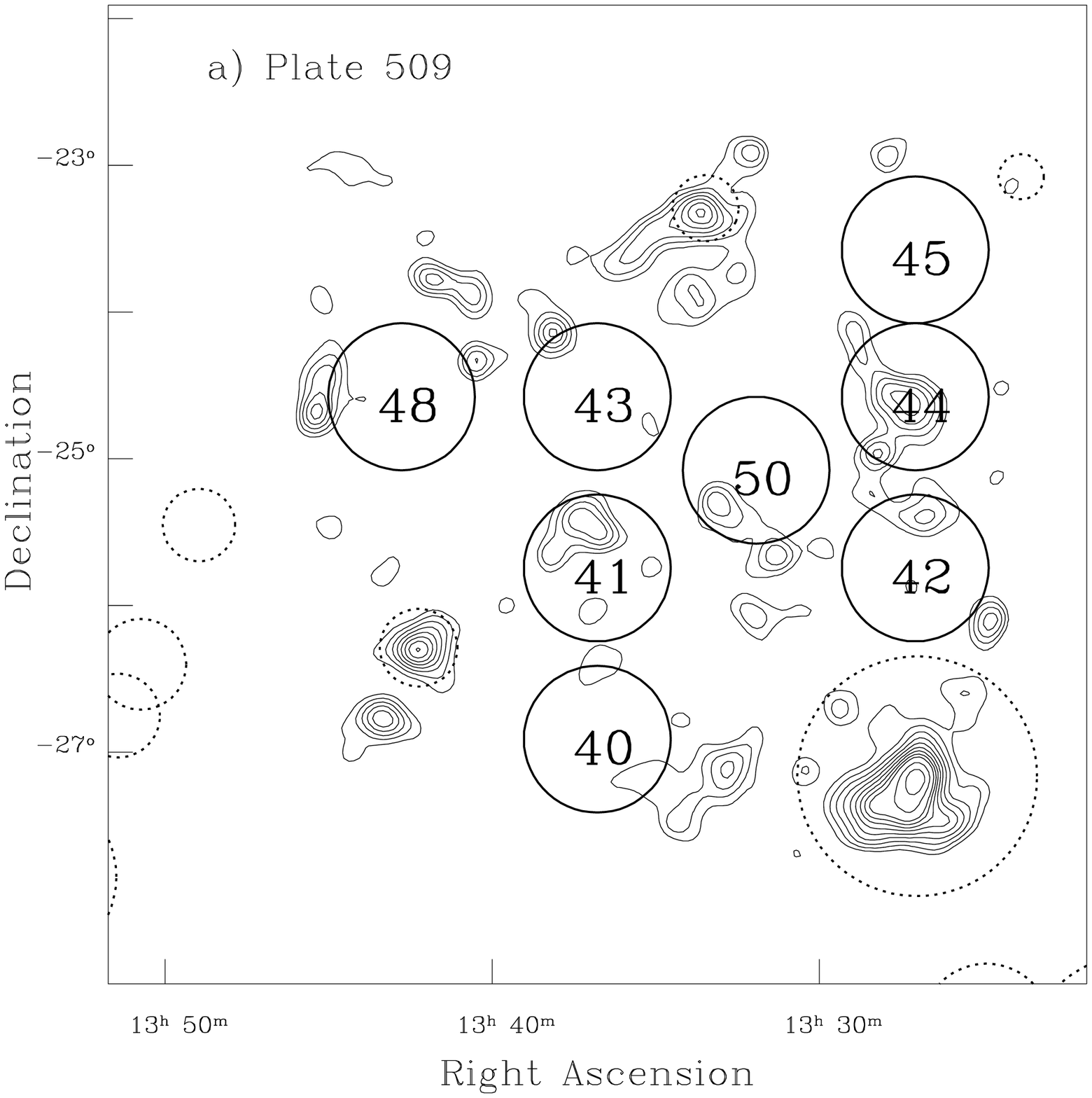} \hfil
\epsfxsize=0.5\hsize \epsfbox{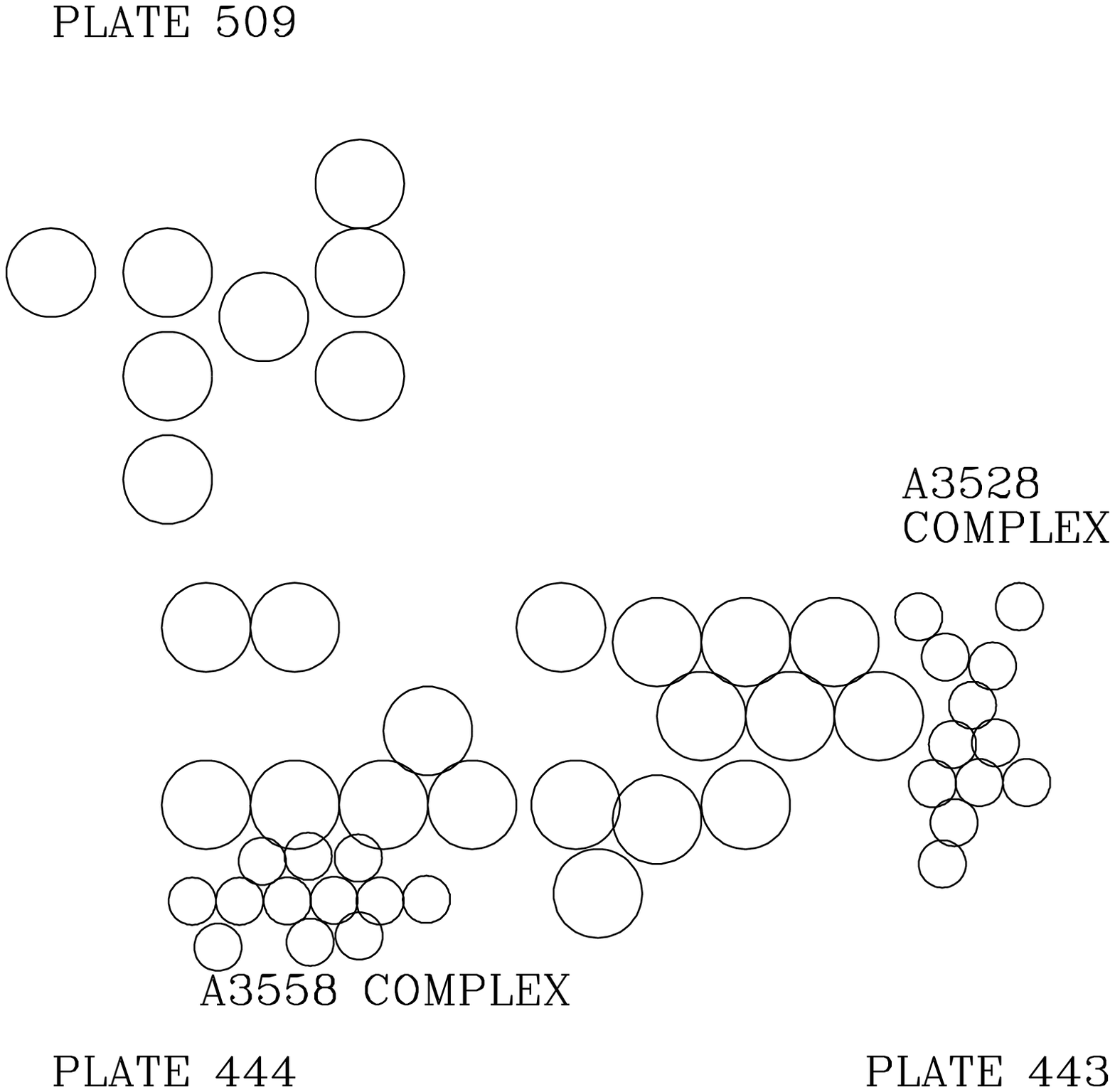} \hfil
\centering
\leavevmode
\epsfysize=8.5cm
\epsfxsize=0.5\hsize \epsfbox{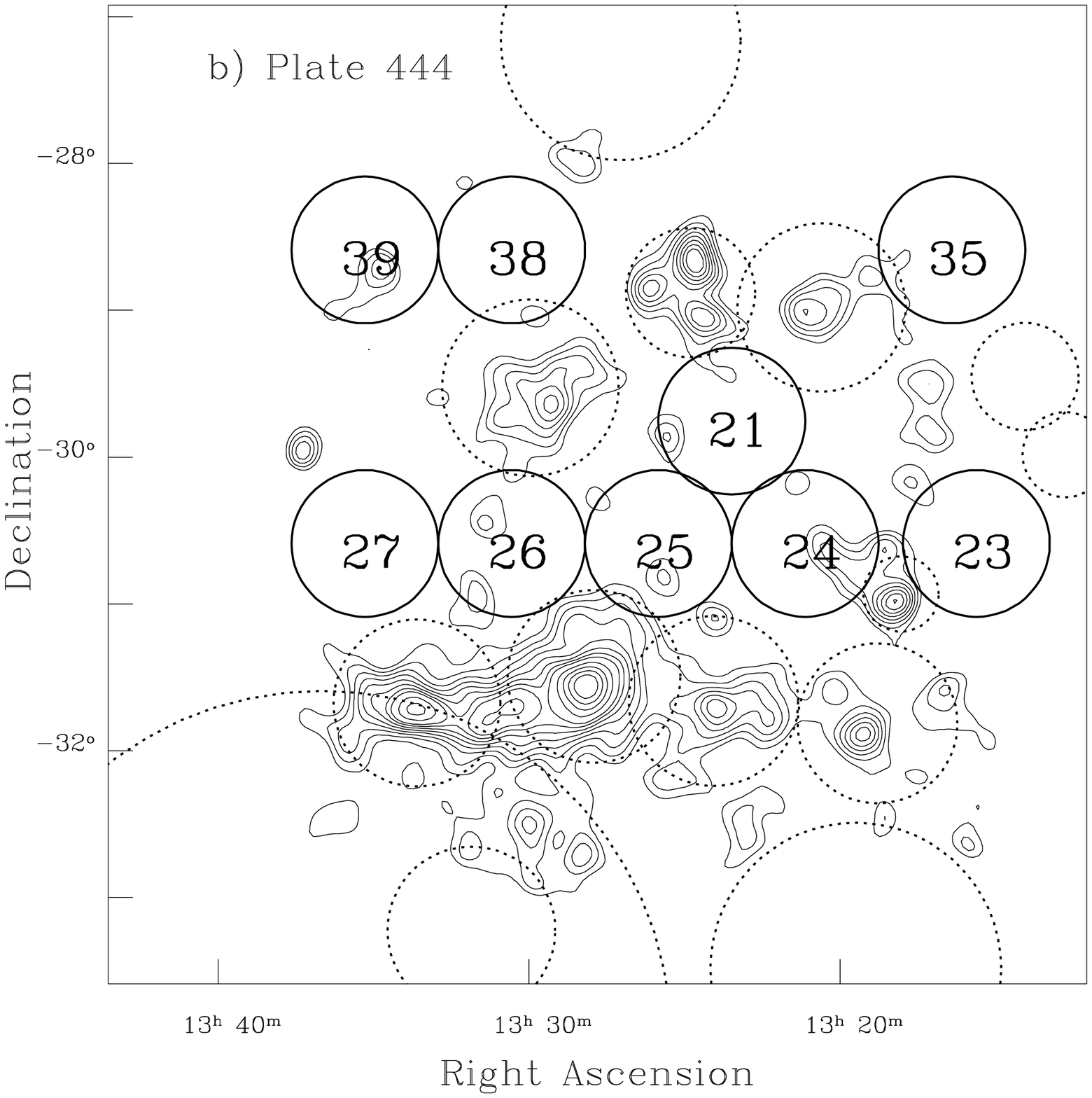} \hfil
\epsfxsize=0.5\hsize \epsfbox{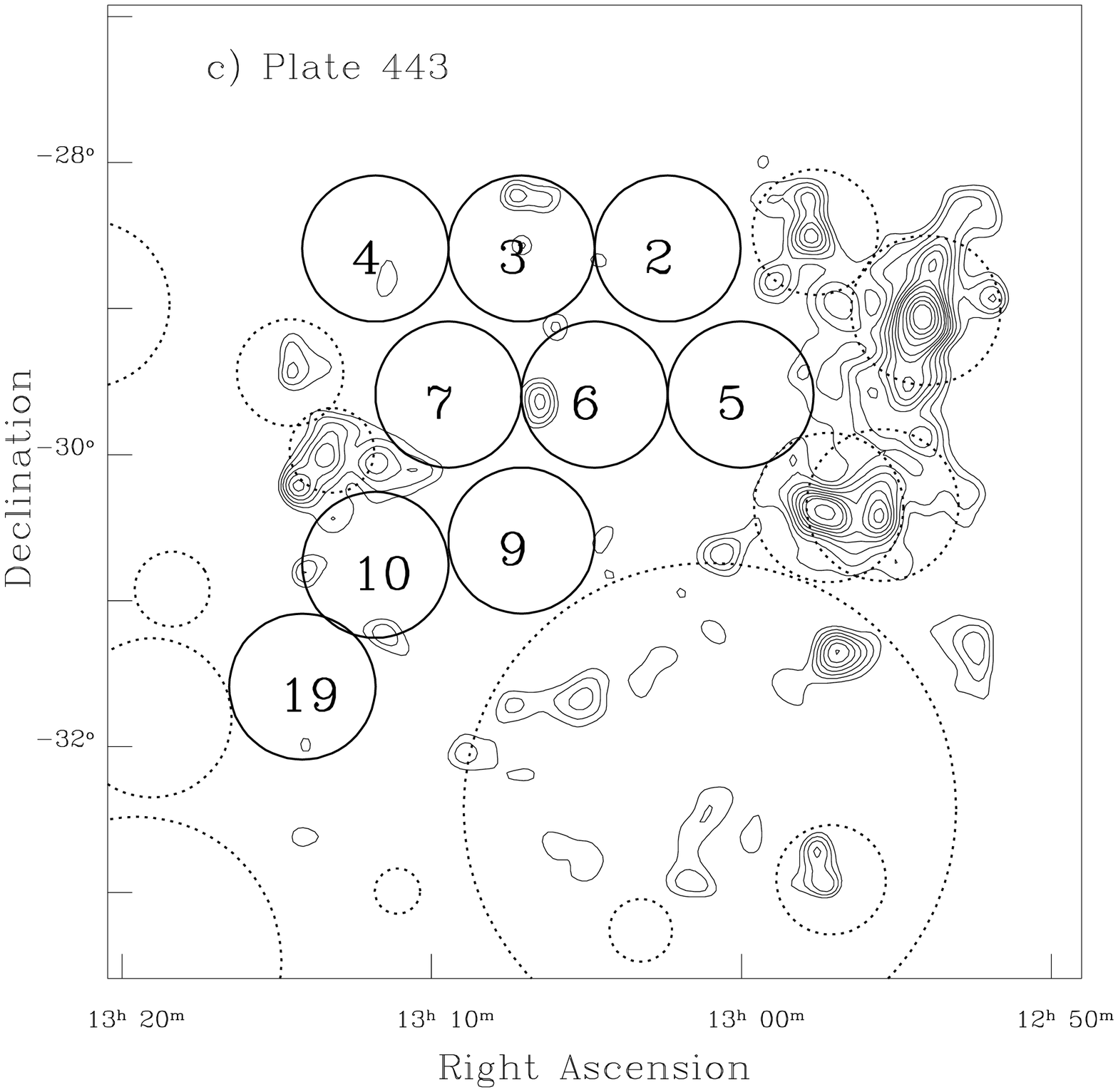} \hfil
\caption[]{Isodensity contours of the bidimensional distribution of the 
galaxies in the $b_J$ magnitude bin $17-19.5$ in the UKSTJ plates which
cover the central part of the Shapley Concentration.
The data have been binned in $2$ $\times$ $2$ arcmin 
pixels and then smoothed with a Gaussian with a FWHM of $6$ arcmin. For the
Abell clusters present in the plates, circles of one Abell radius have been
drawn (dashed curves). Solid circles represent the MEFOS fields. 
\\
a) Plate 509; b) Plate 444; c) Plate 443. The two most evident systems are 
the A3558 (plate 444) and the A3528 (plate 443) cluster complexes. The dashed 
circle at the bottom right corner of plate 509 corresponds to the cluster 
A1736.
\\
The relative positions of the fields is shown in the upper right panel, where
small circles represent observations on the cluster complexes.}
\label{fig:plates}
\end{figure*}

%
%
\section{The intercluster galaxy sample}

As pointed out above, the Shapley Concentration is very rich in clusters
with respect to the other known structures. The aim of the intercluster
survey is therefore to study how the galaxies trace this supercluster and 
which percentage of the mass of the supercluster lies outside the clusters.

\subsection{The catalogue}

Figures \ref{fig:plates}a, b, c show the isodensity contours for the galaxies 
in the $b_J$ magnitude range $17-19.5$ from the COSMOS/UKST galaxy catalogue 
(Yentis et al. 1992) in the three plates 443, 444 and 509, which cover the 
central part of the Shapley Concentration.

In order to check the existence of possible zero point photometric errors in
our data, we compared the COSMOS magnitudes with the CCD sequences of Cunow
et al. (1997), available for plates 443 and 444. 
After having applied the correction for the non-linearity of the COSMOS 
magnitude scale proposed by Lumsden et al. (1997), we found $<b_J - B_{CCD}> 
\sim -0.02 \pm 0.06$ mag using 12 galaxies in the $b_J$ range $16-18.4$. 
Since no CCD sequences are available for plate 509,
we checked its photometric zero point using the galaxies in the region
overlapping the adjacent plate 444, finding that the two magnitude scales
are in agreement. 
We also checked the photometric zero point for ESP survey galaxies 
(Vettolani et al. 1997), which represent our ``field" normalization (see
Sect.5.1), using the CCD sequences of Cunow (1993). We found, using 13
galaxies, $<b_J - B_{CCD}>\sim -0.03 \pm 0.09$ mag. 
These results show, on one side, that there is no systematic zero point
shift in the photometric scale of our plates; on the other side, they confirm
the consistency of our data with respect to the ESP data, which we will use
as normalization in the overdensity estimates (see Sect.5.1).
\\
Finally, we found that a color correction for the conversion from $b_J$ to 
$B_{CCD}$ magnitudes is not required: this fact was already noted
analysing independent CCD sequences on ESP galaxies (Garilli et al., in
preparation).  

The data in Figure \ref{fig:plates} have been binned in $2$ $\times$ $2$ arcmin 
pixels and then smoothed with a Gaussian with a FWHM of $6$ arcmin. For the
Abell clusters present in the plates, circles of one Abell radius have been
drawn (dashed circles). 
The magnitude range of galaxies in the figure has been chosen in order to 
enhance the features at distances equal to or greater than that of the Shapley 
Concentration. In fact, at $14500$ km/s the apparent magnitude of an $M^*$ 
galaxy is $\sim 16.2$.

The solid circles in Figure \ref{fig:plates} correspond to the field of view 
of the multifiber spectrograph MEFOS (one degree of diameter, 
$\sim 0.785$ square degrees), which carries the fibers on rigid arms and 
allows the contemporary observation of 
29 galaxies. This number would match the galaxy density in the $b_J$ magnitude
range $17-18.2$: unfortunately, due to the constraints which avoid arm 
collisions, only part of the objects are observable. Therefore, in order not
to waste fibers, we have chosen to extend the magnitude range to $17.0-18.8$: 
in this range, the average number of galaxies per field is $\sim 70$. 
 
In Table \ref{tab:fields} we report the coordinates of the centre and the 
relative UKSTJ plate of the 26 observed fields. Note that 
in choosing the pointing directions, we avoided clusters, in 
order to observe mainly true ``field" objects, to which hereafter we refer as 
intercluster galaxies.

In the upper right panel of Figure \ref{fig:plates} the relative positions
of the observed fields are shown, together with the positions (small circles)
of the observed fields on the cluster complexes (see Sect. 3). 

\begin{table}
\caption[]{ Centres of the observed MEFOS fields }
\begin{flushleft}
\begin{tabular}{rrrr}
\hline\noalign{\smallskip}
Field & $\alpha$ (2000) & $\delta$ (2000) & $\#$ Plate \\
\noalign{\smallskip}
\hline\noalign{\smallskip}
  2  & 13 02 27 & -28 36 00 & 443 \\
  3  & 13 07 00 & -28 36 00 & 443 \\
  4  & 13 11 38 & -28 36 00 & 443 \\
  5  & 13 00 07 & -29 36 00 & 443 \\
  6  & 13 04 44 & -29 36 00 & 443 \\
  7  & 13 09 21 & -29 36 00 & 443 \\
  9  & 13 07 04 & -30 36 00 & 443 \\
 10  & 13 11 43 & -30 46 00 & 443 \\
 19  & 13 14 04 & -31 36 00 & 443 \\
 21  & 13 23 34 & -29 45 21 & 444 \\
 23  & 13 15 49 & -30 35 21 & 444 \\
 24  & 13 21 14 & -30 35 21 & 444 \\
 25  & 13 25 52 & -30 35 21 & 444 \\
 26  & 13 30 30 & -30 35 21 & 444 \\
 27  & 13 35 09 & -30 35 21 & 444 \\
 35  & 13 16 42 & -28 35 21 & 444 \\
 38  & 13 30 29 & -28 35 21 & 444 \\
 39  & 13 35 04 & -28 35 21 & 444 \\
 40  & 13 36 47 & -26 54 37 & 509 \\
 41  & 13 36 47 & -25 44 37 & 509 \\
 42  & 13 27 11 & -25 44 37 & 509 \\
 43  & 13 36 47 & -24 34 37 & 509 \\
 44  & 13 27 14 & -24 34 37 & 509 \\
 45  & 13 27 16 & -23 34 37 & 509 \\
 48  & 13 42 40 & -24 34 37 & 509 \\
 50  & 13 32 00 & -25 04 37 & 509 \\
\noalign{\smallskip}
\hline
\end{tabular}
\end{flushleft}
\label{tab:fields}
\end{table}

\begin{figure*}
\epsfysize=11cm
\epsfbox{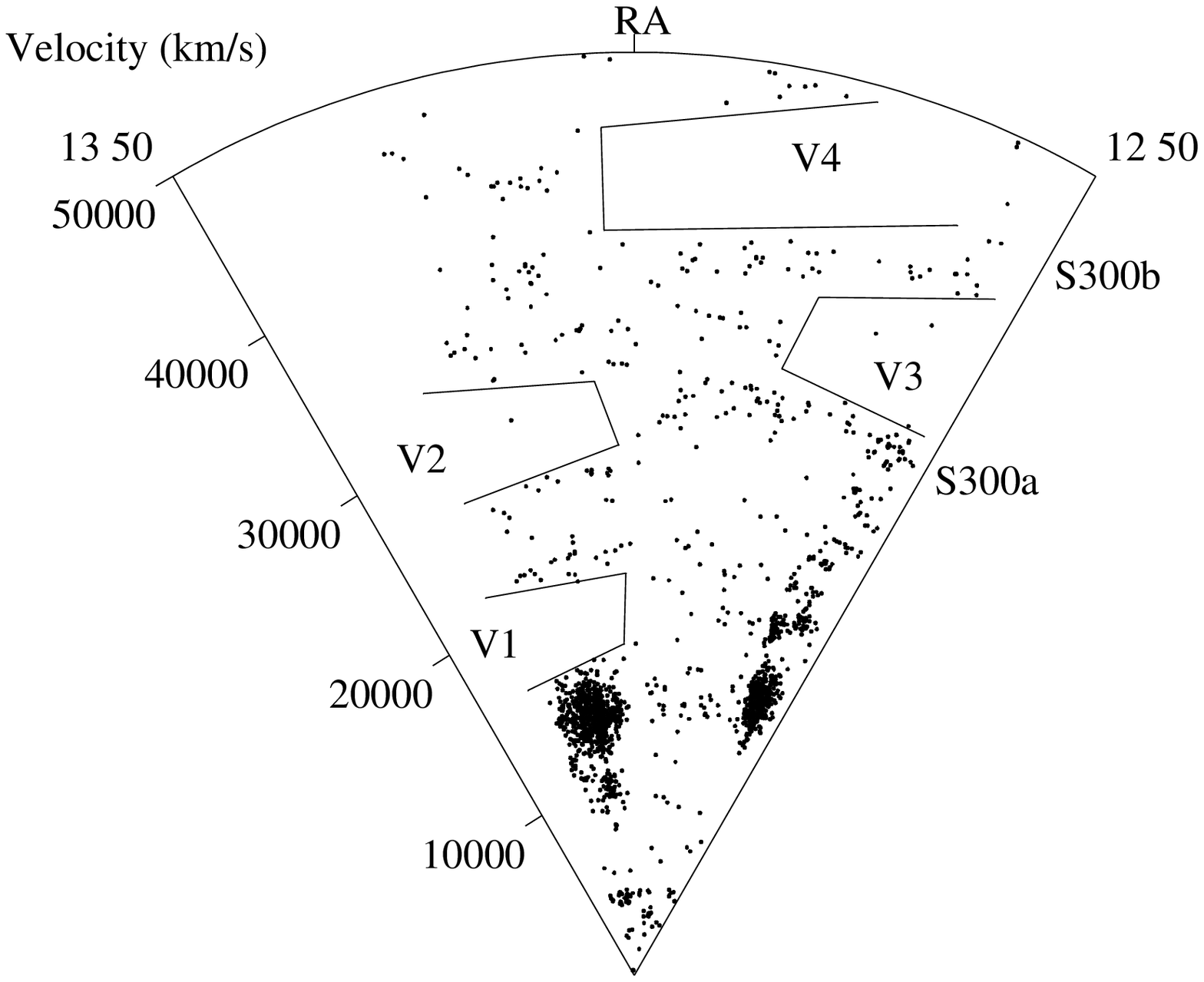}
\epsfysize=11cm
\epsfbox{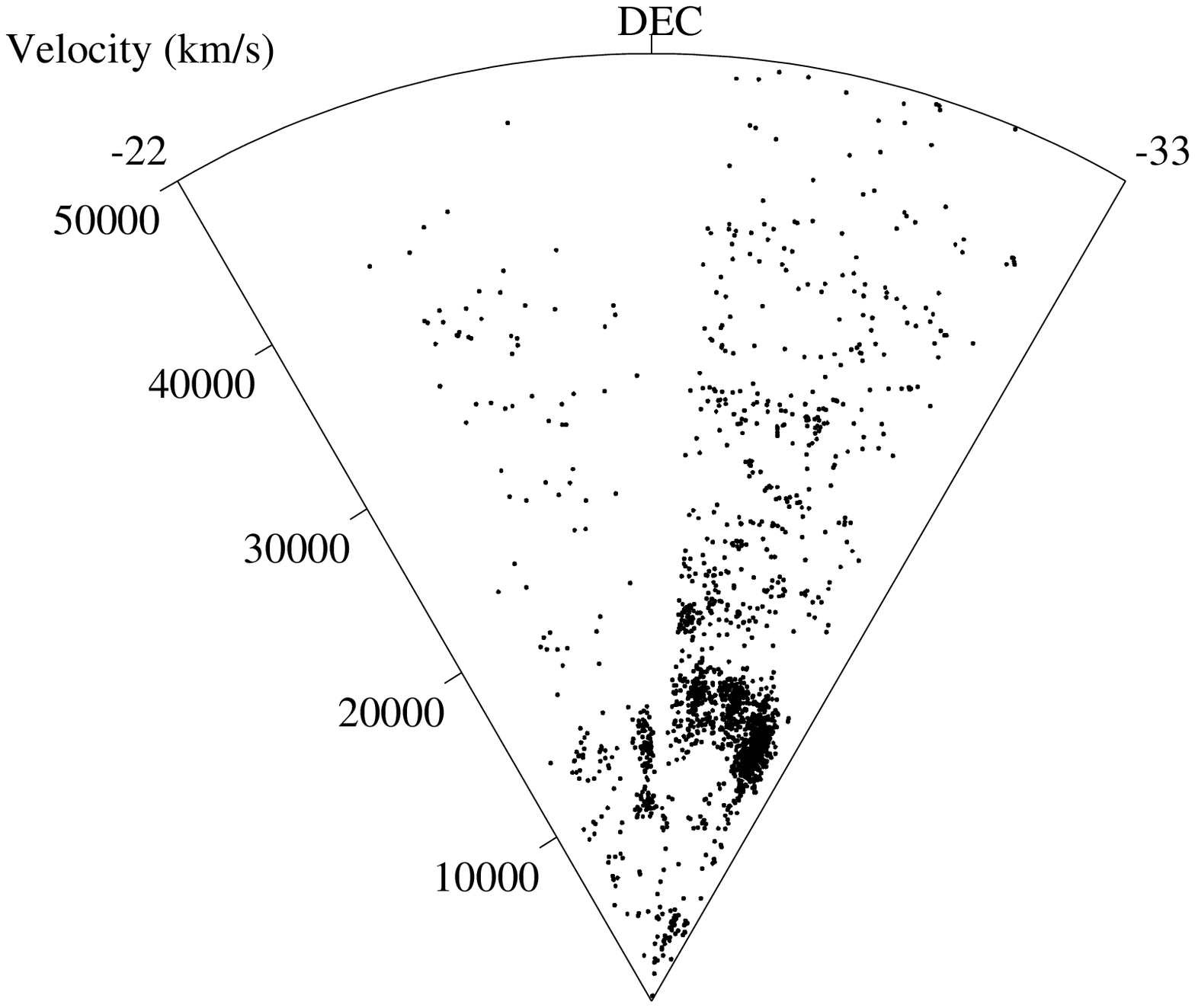}
\caption[]{Wedge diagram of the galaxy distribution in our samples.
Upper panel: 1728 galaxies in the plates 444 and 443 and in the velocity 
range $[0-50000]$ km/s. Note the clear presence of the A3558 and A3528 
complexes and of four voids (see text for details). 
Lower panel: 1979 galaxies in plates 443, 444 and 509 in the velocity range 
[0-50000] km/s. Note that the plotted coordinate is the declination.}
\label{fig:wedge}
\end{figure*}

\subsection{Observations and data reduction}

Spectroscopic observations were performed at the 3.6m ESO telescope at La Silla,
equipped with the MEFOS multifiber spectrograph (Bellenger et al. 1991; 
Felenbok et al. 1997) in the
nights of 8-9-10 April 1994. The MEFOS multifiber spectrograph was mounted
at the prime focus of the telescope and allowed the contemporary observation 
of $29$ scientific targets. Each fiber, whose projected diameter on the sky is
$\sim 2.6$ arcsec, was coupled with another one devoted to
a simultaneous sky observation. This has in principle two advantages:
the sky spectrum is measured very nearby the considered galaxy and the
exposure is taken at the same time. 
Moreover, after having checked that background spatial variations are not 
present, it is possible to average the 29 sky spectra in order to obtain a 
``mean" sky with an enhanced signal--to--noise ratio. 

As pointed out in Bardelli et al. (1994), the sky subtraction procedure
in fiber spectroscopy requires the determination of the relative fiber 
transmissions, which we estimated on the basis of the [OI]$\lambda$ 5577
sky line. 
MEFOS allowed also the possibility of ``beam-switching", i.e. the change between
the object and the nearby sky fiber. Dividing the observation in two 
exposures and applying this option, galaxy and sky fall on the same fiber in 
different moments: this would allow a direct subtraction of the sky from the
galaxy spectrum. However, even if this procedure avoids the problems given by 
the fiber-to-fiber different transmission, it can fail to correctly subtract 
the sky when the latter changes significantly between the two exposures. 
We found that this happened at large zenithal angles or at the beginning and
at the end of the night. However, we used both methods and we always considered 
the procedure which gives the higher number of measured radial velocities 
in each field.

The spectra, obtained with a CCD TEK512 CB and the ESO grating $\# 15$,
have a resolution of $\sim 12$ \AA~and a pixel size of $\sim 4.6$ \AA.
They were cross--correlated with a set of 8 stellar and 8 galaxy
templates using the XCSAO program of Kurtz et al. (1992). For spectra
showing emission lines, we used the EMSAO program in the same package.
Details of the reductions and the cross-correlation are given in Bardelli
et al. (1994). 

\subsection{ The redshift sample}

We observed a total of 685 objects, instead of the maximum number of 754 
in principle allowed by the number of 
fibers (29 $\times$ 26 fields), due to collision constraints in the robotic
arms of the MEFOS spectrograph. Among these 685 spectra, 85 were not useful
for a redshift determination ($12\%$ of the total), due to a poor signal to 
noise ratio. Among the 600 good spectra, 158 objects resulted to be stars
($26\%$ of the total), leaving us with 442 galaxy velocities. 

In Table \ref{tab:sample}, we list the galaxies with velocity determination. 
Columns (1), (2) and (3) give the 
right ascension (J2000), the declination (J2000) and the $b_J$
apparent magnitude of the object, respectively. Columns (4) and (5) give the
heliocentric velocity ($v=cz$) and the internal cross correlation error: 
this value has to be moltiplied by a factor 1.6-1.9 in order
to find the true statistical error (see Bardelli et al. 1994). The code 
``emiss" in column (6) denotes the velocities obtained from emission lines.

Finally in the following analysis we included also 69 velocities found in the 
literature (not reported in Table 
\ref{tab:sample}) for objects in our surveyed region and therefore our total 
sample has 511 velocities. The average redshift completeness of the sample 
in the magnitude range $17-18.8$ is $\sim 25 \%$.

%
%
\section{The cluster galaxy sample}

\subsection{The A3558 cluster complex}

Figure \ref{fig:plates}b clearly shows the presence on the plate 444 of an 
elongated stucture 
formed by the ACO clusters A3558, A3562 and A3556, which is located at the
geometrical centre of the Shapley Concentration. This complex has been
extensively studied in the optical
(Bardelli et al. 1994, 1998a, 1998b), in the X-ray 
(Bardelli et al. 1996) and in the radio (Venturi et al. 1997, 1998)
wavelengths. 
We found that this structure may be a cluster-cluster collision seen just
after the first core-core encounter. 
In order to determine the excess in number of galaxies, we used the redshift
sample presented in Bardelli et al. (1998a), consisting of 714 velocities
over an area of $3^o.12 \times 1^o.4$,
to which we added the 60 new redshifts published by the ENACS survey 
(Katgert et al. 1998).
In order to maximize the completeness, we restricted our analysis to the
well sampled region corresponding to the OPTOPUS fields (see Bardelli et al.
1998a), covering an area of about $2.7$ square degrees.
In this sample, there are 723 galaxies with velocity determination, out of a
total of 1582 ($\sim 46 \%$) to the limit $b_J=19.5$. Restricting the sample in
the magnitude range $17-18.8$, the completeness is $\sim 64\%$ (456/711). 

\subsection{The A3528 cluster complex}

Another remarkable cluster complex is found at the westernmost part of plate 
443 (Figure \ref{fig:plates}c) and is formed by the ACO clusters A3528, A3530
and A3532. We performed a redshift survey in this structure with the 
OPTOPUS multifibre spectrograph, obtaining 581 new velocities in an area of 
$3^o.38 \times 1^o.72$, including also the cluster A3535 (Bardelli et al., 
in preparation). After a search in the literature, we found 
79 additional velocities from Quintana et al. (1995) and ENACS (Katgert
et al. 1998), leading to a final sample of 660 redshifts. As done
for the A3558 complex, we restricted the sample to the area sampled
by the 12 OPTOPUS fields (about $2.7$ square degrees). In this area 
the completeness is $\sim 46\%$ (645/1399) to $b_J=19.5$ and $\sim 72\% $ 
(475/656) in the magnitude range $17-18.8$.

\subsection{Other clusters}

In the region under consideration, in addition to the clusters cited above,  
there are several other Abell/ACO clusters.
Among them, A3537 and A3565 are part of the Great Attractor, while
A1736, A3554, A3555, A3559 and A3560 are members of the main
structure of the Shapley Concentration. The cluster A3552, westward of A3556,
was not previously included in this supercluster by Zucca et al. (1993) because
of the value of its estimated redshift: now the available redshifts for this
cluster (Quintana et al. 1995) put it in the distance range of the Shapley
Concentration. An ambiguous case is that of A3557, which has two velocity peaks:
only one of them is compatible with the supercluster.
\\
Among the remaining clusters, A1757, A1771, A3540, A3546 and A3551 may 
belong to another structure at $30000$ km/s (S300, see below). 
Finally the clusters A3531 and A3549 are at intermediate distance between 
Shapley Concentration and S300, while A1727 and A3544 are more distant objects, 
on the basis of their estimated redshifts.
\\ 
Among the clusters which are part of the Shapley Concentration,
the only one which has been sampled enough to be useful in our analysis is 
A1736, clearly visible in the lower right corner of plate 509 (Figure 
\ref{fig:plates}a). The redshifts in A1736 have been 
taken by Dressler \& Shectmann (1988) and Stein (1996), and the final sample 
contains 111 velocities (54 in the magnitude range $17-18.8$).
For the others, the redshift data found in the literature are too sparse 
to allow density reconstructions, therefore we choose to neglect these clusters.
We ignore also the more distant clusters, because of the lack of data.
Therefore, all the following results for the Shapley Concentration and S300 
have to be regarded as {\it lower limits} for what concerns the cluster 
contribution.

\begin{figure*}
\centering
\leavevmode
\epsfysize=8.5cm
\epsfxsize=0.5\hsize \epsfbox{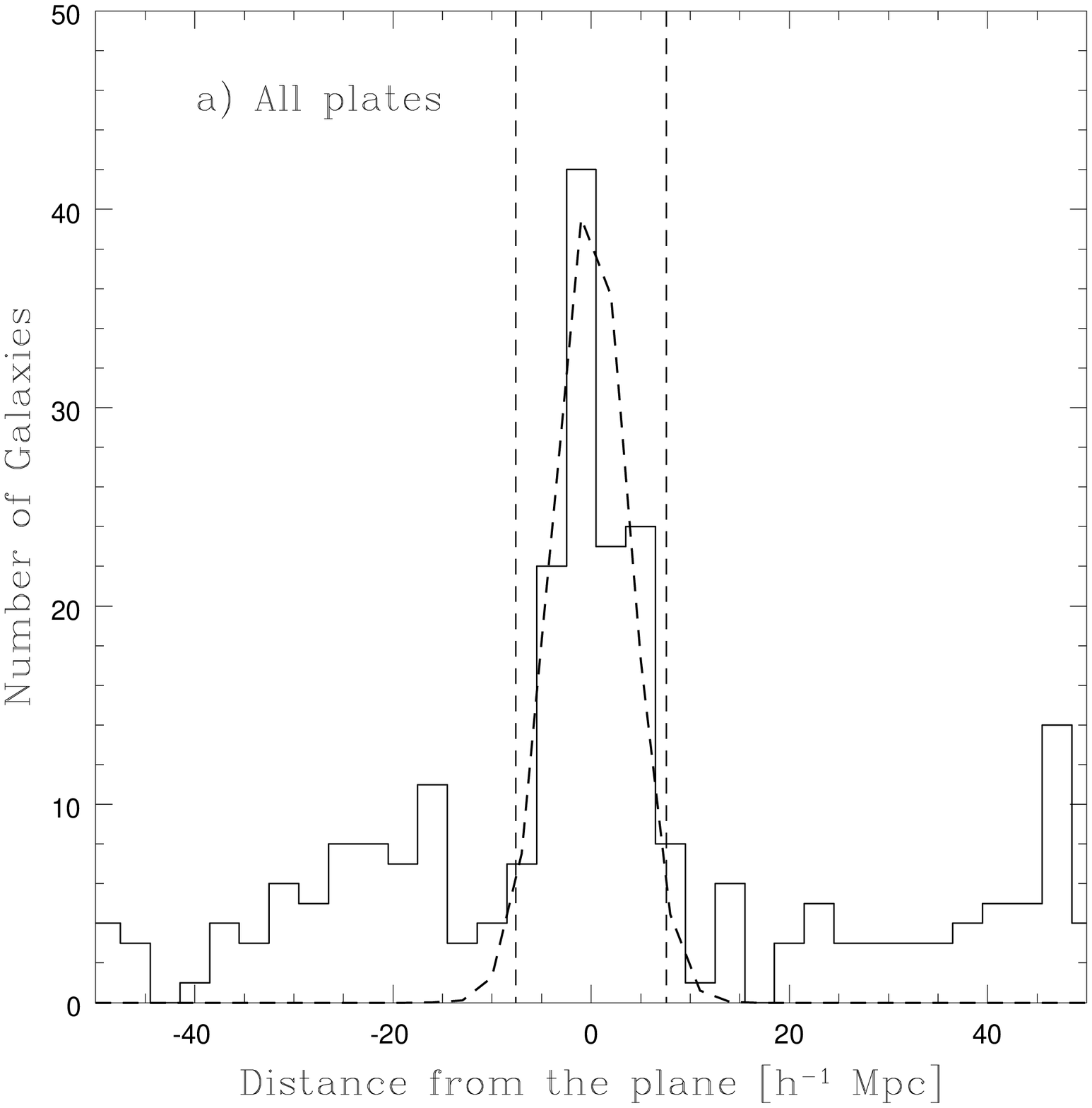} \hfil
\epsfxsize=0.5\hsize \epsfbox{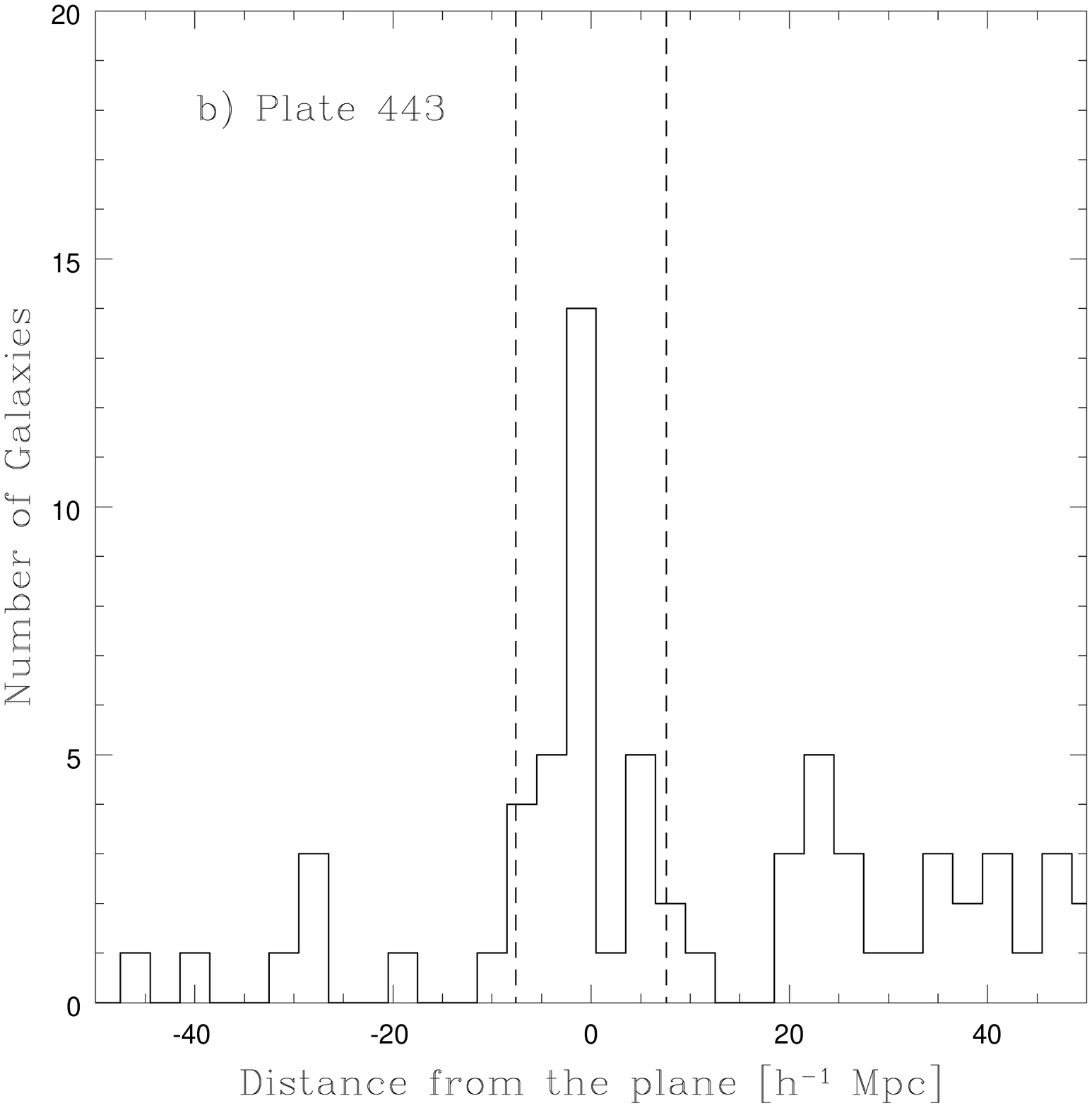} \hfil
\centering
\leavevmode
\epsfysize=8.5cm
\epsfxsize=0.5\hsize \epsfbox{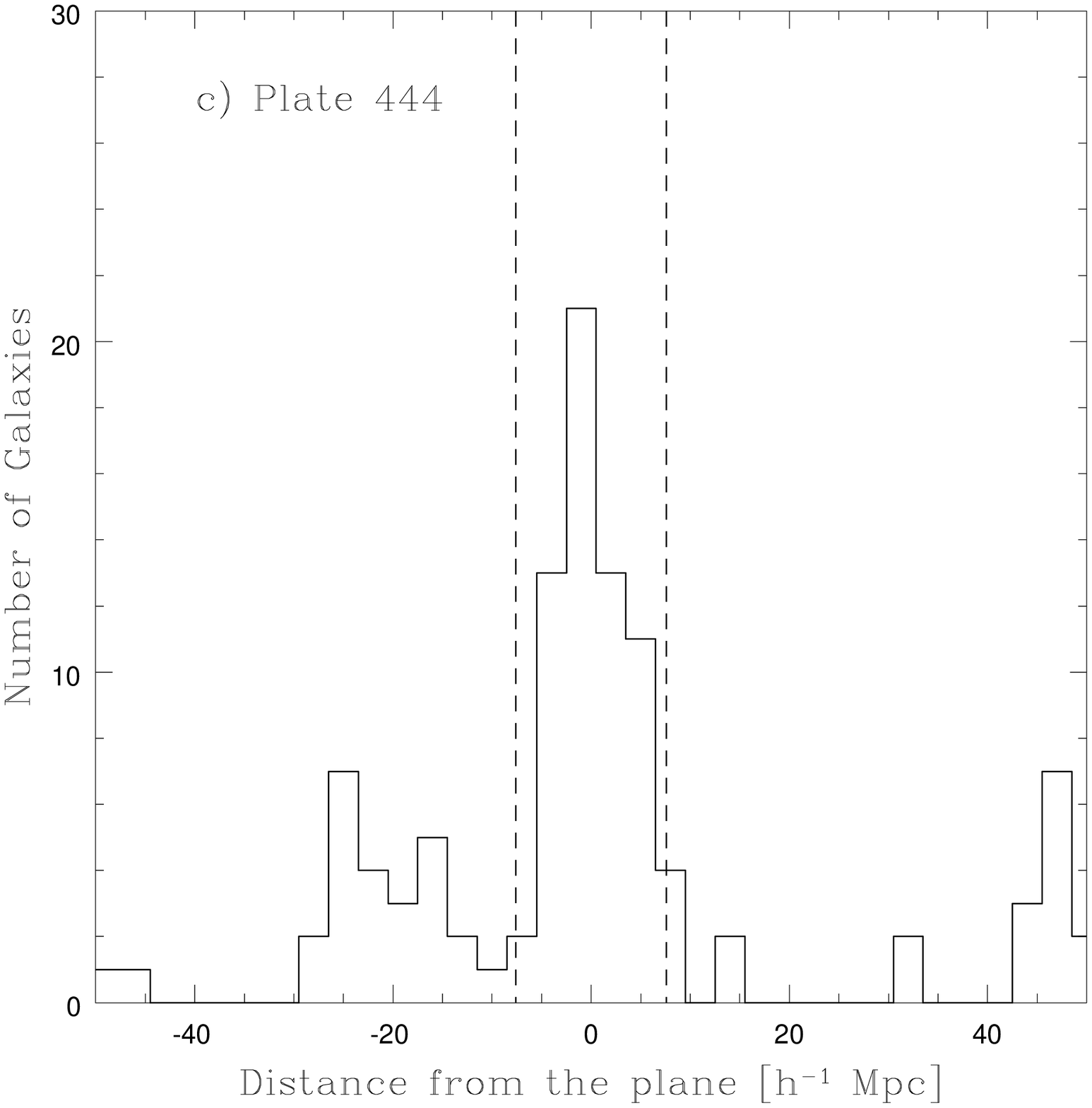} \hfil
\epsfxsize=0.5\hsize \epsfbox{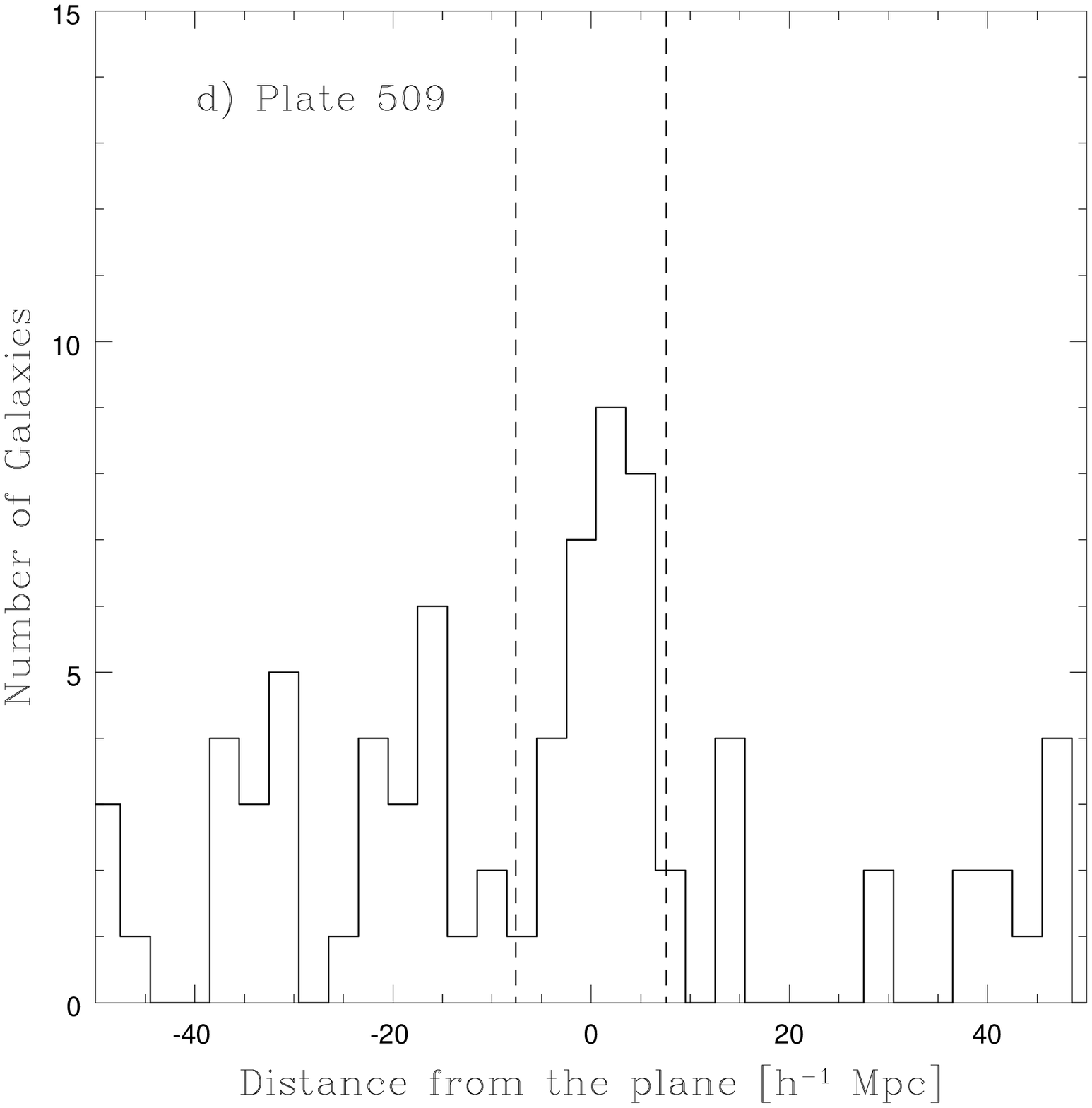} \hfil
\caption[]{ Histograms of the distances from the fitted plane 
for all galaxies in the intercluster survey (panel a) and for galaxies in 
each plate (panels b, c, d). The vertical dashed lines indicate
the interval assumed for the estimate of the supercluster 
overdensity. In the upper left panel the fitted Gaussian with 
$\sigma=3.8$ \hmpc is drawn. } 
\label{fig:distplane}
\end{figure*}

%
%
\section{Analysis of the galaxy distribution}

The total velocity sample we used contains 2057 velocities.
In Figure \ref{fig:wedge}a, we plot the wedge diagram of the galaxies with 
redshifts in the plates 444 and 443. The two cluster complexes dominated by 
A3558 (on the left) and by A3528 (on the right) are clearly visible:  
these two structures appear to be connected by a 
bridge of galaxies, similar to the Coma-A1367 system, the central
part of the Great Wall. The scale of this system is $\sim 23$ \hmpc and it is
comparable to that of Coma-A1367 ($\sim 21$ \hmpc). Note the presence
of two voids at $\sim 20000$ km/s and $\sim 30000$ km/s in the 
easternmost half of the wedge, labelled as V1 and V2 respectively. 
Other two voids (V3 and V4) are visible in the westernmost part of the plot:
in particular, void V3 appears to be delimited by two elongated features 
(S300a and S300b) which appear to ``converge" in a single feature at right 
ascension $\sim 13^h$ (S300c). We refer to this structure with the name S300: 
as shown below, the overdensity corresponding to this excess is remarkably 
high, even if it is not clear if all these features are part of a single
structure.

Figure \ref{fig:wedge}b shows the distribution of all the galaxies 
(plates 443, 444 and 509) of the sample in the velocity range $[0-50000]$
km/s. Note that the plotted sky coordinate is the declination.
The apparent ``hole"  in the galaxy distribution in the middle of the
wedge ($\delta \sim 27^o$) is due to a poor sampling of the southern part of 
plate 509 and to the fact that the literature data for A1736 are less deep
than those in our survey.

\subsection{The structure of the Shapley Concentration }

The average velocity of the observed intercluster galaxies appears to
be a function of the ($\alpha$, $\delta$) position, ranging from 12289 km/s for
the galaxies in plate 509 to 15863 km/s for the galaxies in plate 443. 
This fact was also noted by Quintana et al. (1995) as a relationship between
the cluster radial velocities and the right ascension.
Inspection of the trend of the average velocity as a function of
position suggests that it can be reasonably described by a plane in
the three--dimensional space ($\alpha$, $\delta$, $v$):

\begin{equation}
a(x-x_m) +b(y-y_m) +c(z-z_m) =0 \ ,
\label{eq:plane}
\end{equation}
where 
\begin{eqnarray}
x&=&{v\over{100}} \cos(\delta-\delta_m) \sin(\alpha-\alpha_m)  
\ \ {\rm h^{-1}\ Mpc} \nonumber \\ 
y&=&{v\over{100}} \sin(\delta-\delta_m)             
\ \ {\rm h^{-1}\ Mpc} \nonumber   \\ 
z&=&{v\over{100}} \cos(\delta-\delta_m) \cos(\alpha-\alpha_m) 
\ \ {\rm h^{-1}\ Mpc} \nonumber  
\end{eqnarray}
and the subscript $m$ indicates the average value of the variable. 
\\
Each galaxy has a distance from the plane

\begin{equation}
d={ {a(x-x_m) +b(y-y_m) +c(z-z_m)}\over{\sqrt{a^2+b^2+c^2}} } \ \ 
{\rm h^{-1}\ Mpc.}
\label{eq:distplane}
\end{equation}
Minimizing the sum of the squared distances of galaxies from the plane, 
we find

\begin{equation}
a=0.9\ \ \ b=1.4\ \ \ c=1.1\ .\ \  \nonumber
\end{equation}

In Figure \ref{fig:distplane} we show the histograms of the distances from
the fitted plane. In panel a) the distribution of distances of the 
whole intercluster sample is shown: the peak corresponding to the Shapley 
Concentration is clearly visible. The secondary bump at 
$d \sim -20$ \hmpc corresponds to the structure at $\sim 11000$ km/s
found by Drinkwater et al. (1999). 
This plane, which describes the ridge of the distribution of the galaxies, 
is a reasonably good fit over the entire extension of our survey,
as shown by the distance histograms of galaxies
divided in the three surveyed plates (Figure \ref{fig:distplane}b, c, d). 

\begin{figure}
\epsfysize=8.5cm
\epsfxsize=\hsize \epsfbox{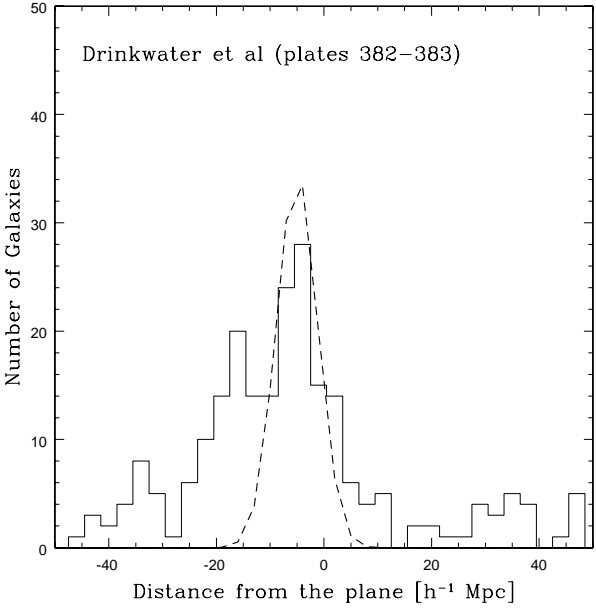} \hfil
\caption[]{Histogram of the distances from the plane of the galaxies
in Plates 382 and 383 from the Drinkwater et al. (1999) survey. Superimposed
is the Gaussian with $\sigma= 3.8$ \hmpc and a mean distance of $-5.0$ \hmpc.}
\label{fig:drinkwater}
\end{figure}

A Gaussian fit to the distribution of the distances around the best fit plane 
gives a dispersion $\sigma =3.8$ \hmpc. This dispersion is significantly 
narrower than what would be obtained ($\sigma_{vel} = 1011$ km/s) by fitting a 
Gaussian to the velocity distribution of the galaxies in the same region.
\\
We checked also if this representation of the galaxy distribution holds outside
our surveyed region using the velocity sample of Drinkwater et al. (1999), which
covers plate 444 and two southern adjacent plates (382 and 383).
Although the plane parametrization seems correct, we find a shift in the mean 
of $\sim -5$ \hmpc in the plates 382 and 383. 
In Figure \ref{fig:drinkwater} we show the histogram of the distances
from the plane of galaxies in plates 382 and 383.
\\
In Figures \ref{fig:distplane}a and \ref{fig:drinkwater}
the fitted Gaussian is superimposed to the histograms. Note that in Figure
\ref{fig:drinkwater} the mean has been shifted by $-5.0$ \hmpc. 

%
\section{Density excesses and masses}

\subsection{The density excess estimate }

In order to estimate the overdensity associated to each structure and to 
reconstruct the density-distance profile, it is necessary to determine the 
number of galaxies expected in the case of a uniform distribution, under the 
same observational constraints (i.e. with the same redshift incompleteness),
and the volume occupied by each structure. 

\begin{figure*}
\centering
\leavevmode
\epsfysize=8.5cm
\epsfxsize=0.5\hsize \epsfbox{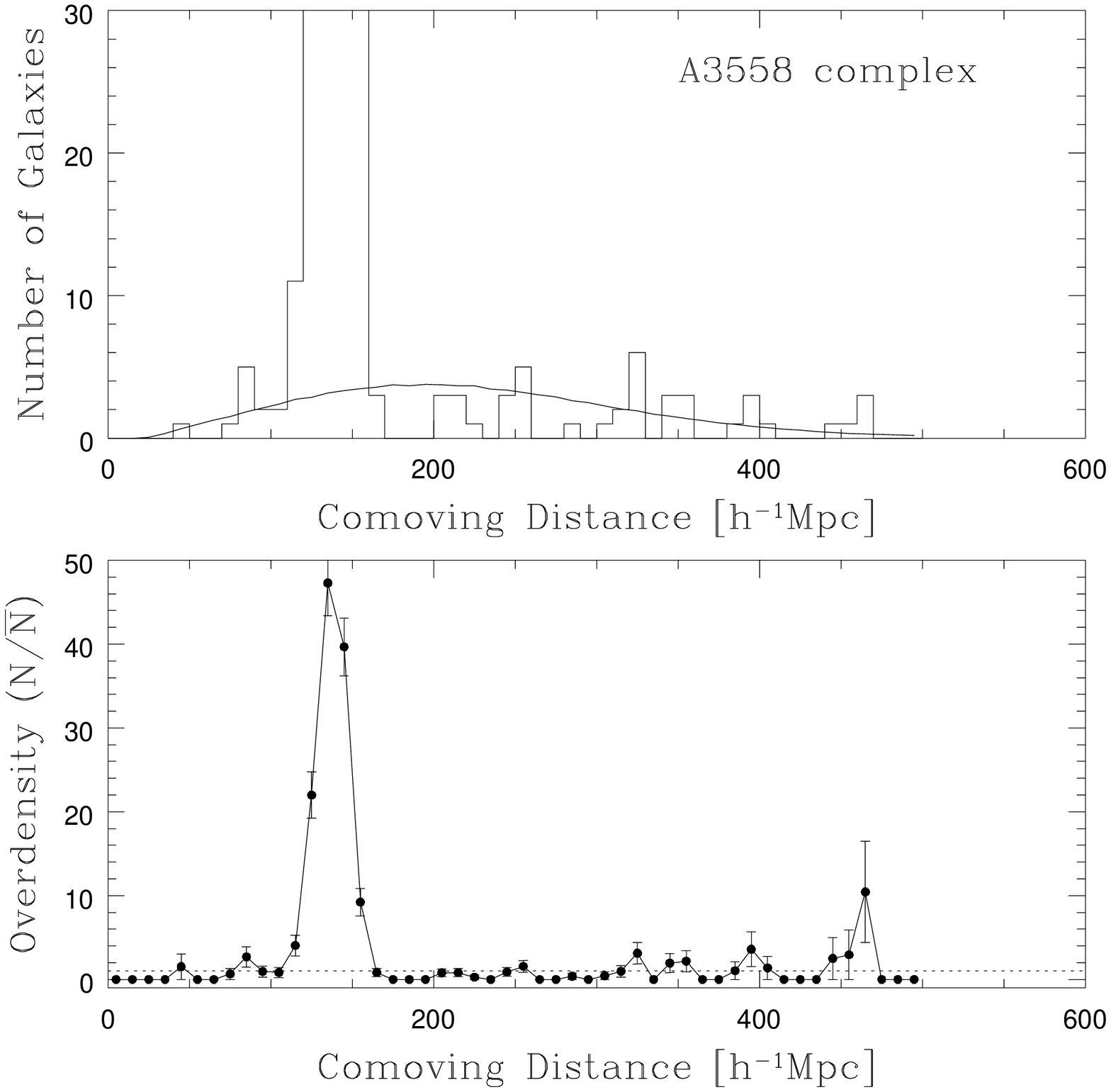} \hfil
\epsfxsize=0.5\hsize \epsfbox{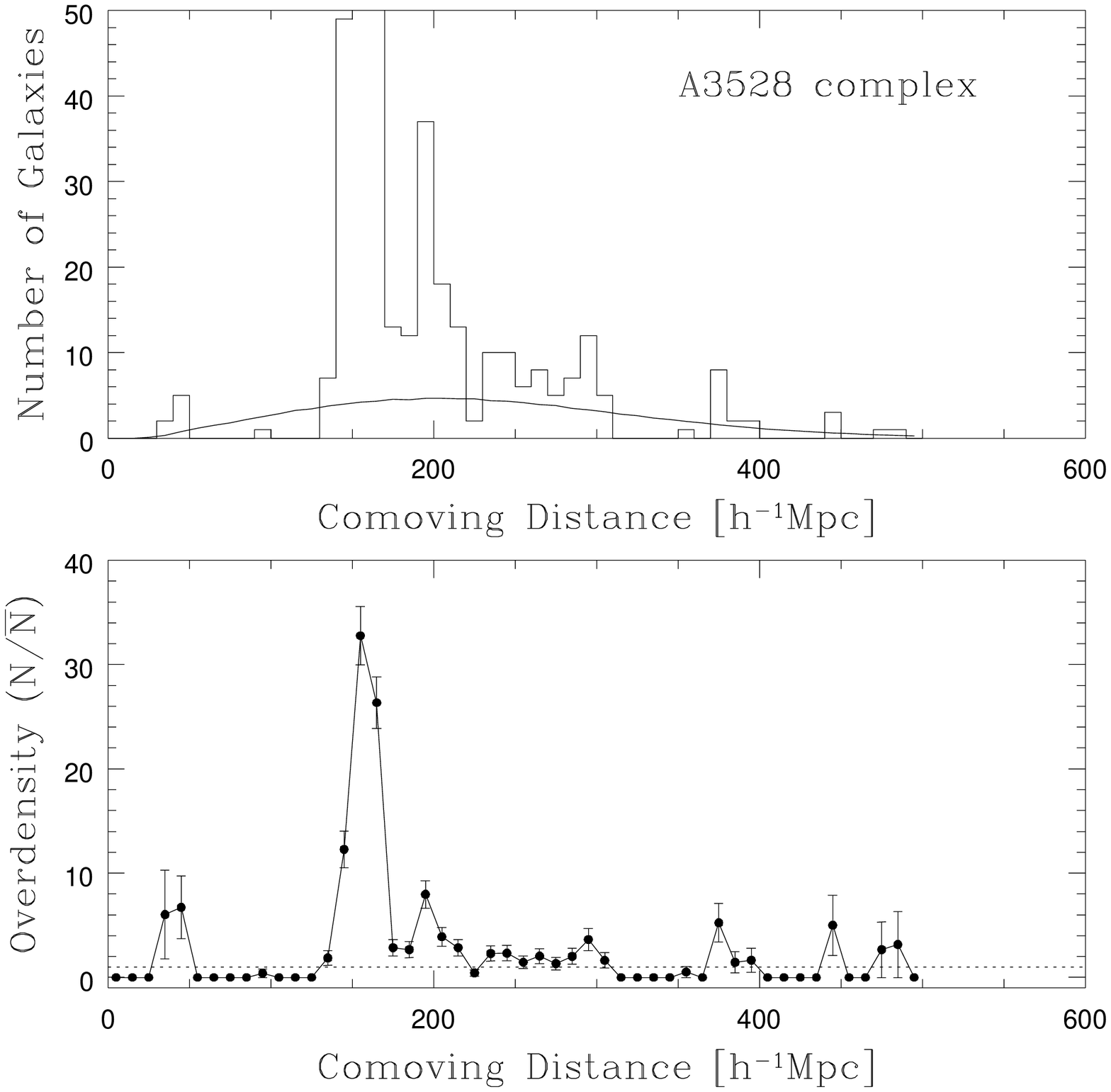} \hfil
\caption[]{Upper panels: Histograms of the velocities in the A3558 (left) and
A3528 (right) complexes with superimposed the distribution expected for a
uniform sample. Note that the histograms have been truncated on the Y-axis in 
order to better show the distribution of field galaxies: in the central peak
there are $\sim 150$ galaxies for A3558 and $\sim 140$ galaxies for A3528.
Lower panels: Galaxy density profiles. The dotted line
corresponds to $\displaystyle{ { N \over \bar{N}}  = 1}$.
The errors are computed assuming Poissonian fluctuations of the number
of detected objects in the considered bin.}
\label{fig:prevcomplex}
\end{figure*}

\begin{figure*}
\centering
\leavevmode
\epsfysize=8.5cm
\epsfxsize=0.5\hsize \epsfbox{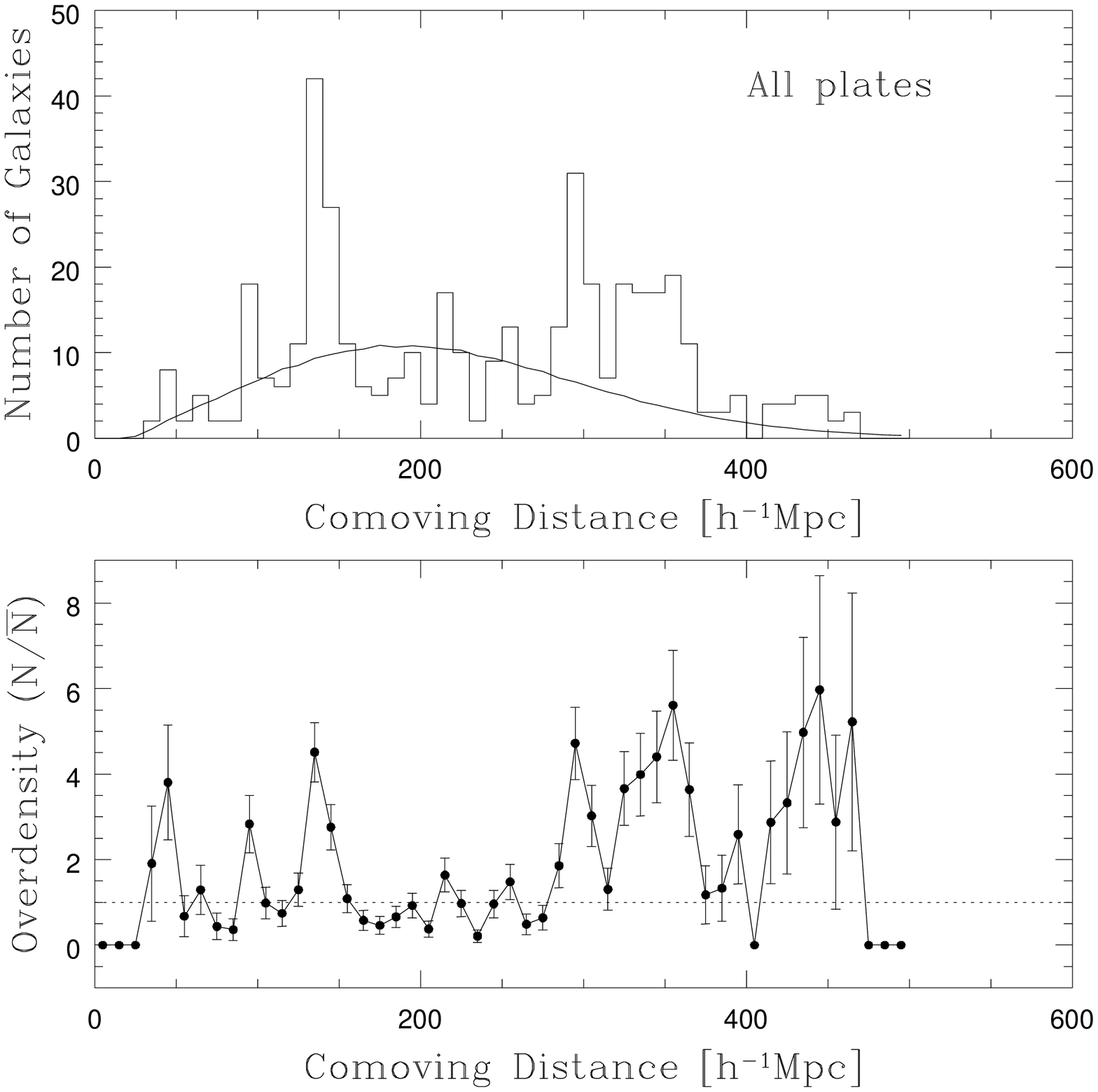} \hfil
\epsfxsize=0.5\hsize \epsfbox{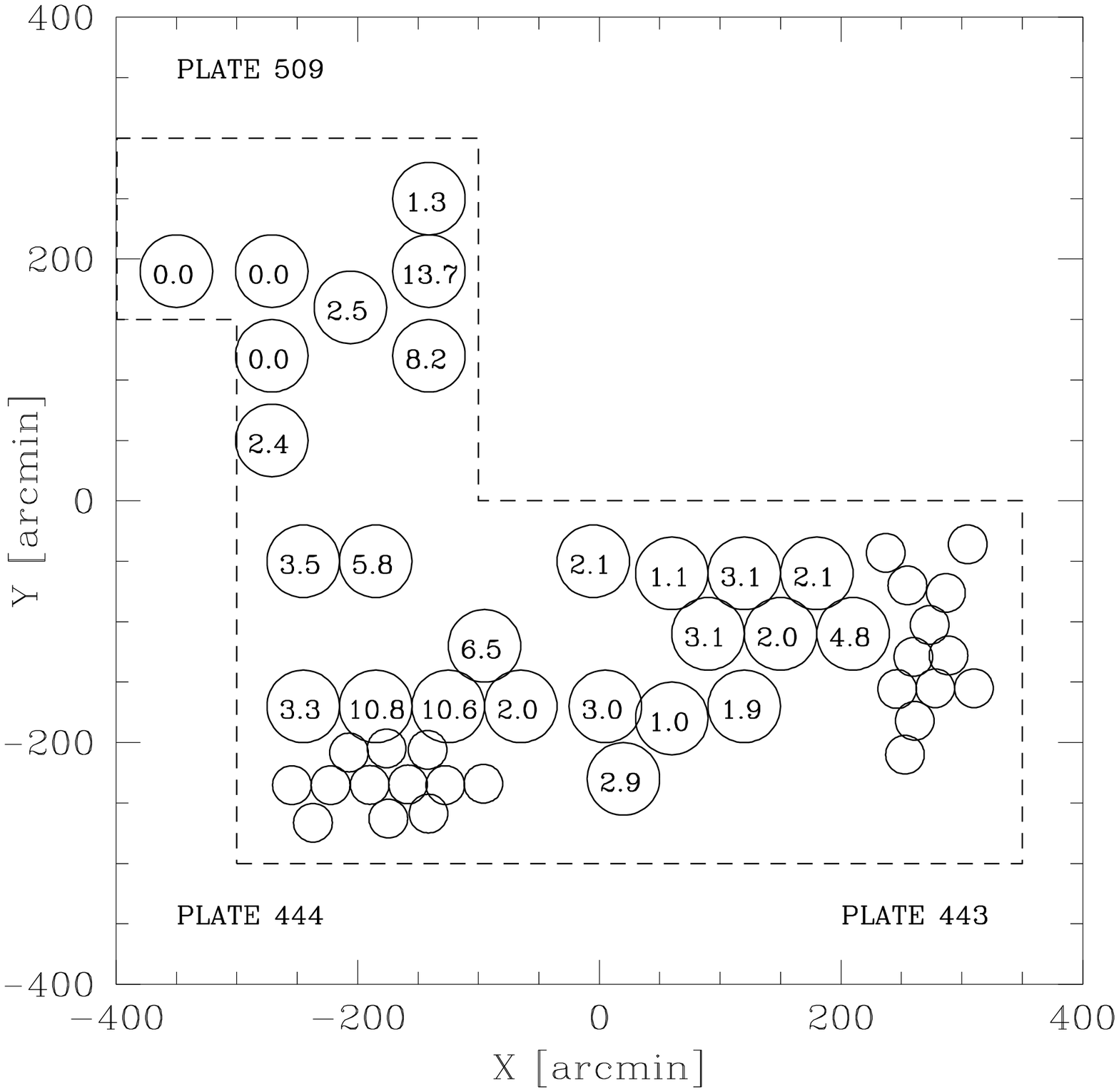} \hfil
\caption[]{Left panel: Same as Figure \ref{fig:prevcomplex} for the intercluster
galaxies sample. Right panel: estimated overdensities in the single MEFOS 
fields as distributed in the sky. }
\label{fig:prevplates}
\end{figure*}

The first step is the determination of the selection function of the
survey.
For each sample, we computed the expected number of galaxies in the case
of uniform distribution as

\begin{equation}
N(z) dz = \sum_i C(m_i) \int^{L_{max}(m_i,z)}_{L_{min}(m_i,z)}  \phi(L)   
{ { dV}\over{dz}}  dL dz \ ,
\end{equation}
where $V$ is the volume, $L_{min}(m_i,z)$ and $L_{max}(m_i,z)$ are the minimum 
and maximum luminosity seen in the sample at the redshift $z$, given the 
apparent magnitude limits of the sample. The quantity $C(m_i)$ represents
the incompleteness of the 
survey, i.e. the ratio between the number of galaxies with redshifts and the 
total number of objects, as a function of the apparent magnitude, and 

\begin{equation}
\phi (L)dL= \phi^* \left( {{L}\over{L^*}}\right)^{\alpha} e^{-L/L^*} 
d \left({{L}\over{L^*}}\right) 
\end{equation}
is the luminosity function of field galaxies parametrized with a Schechter 
(1976) function. We adopted the values 
$\alpha=-1.22$, $M^*=-19.61$ and $\phi^*=0.02$ found by Zucca et al. (1997)
for the ESP survey, which uses our same photometric catalogue.
We applied the correction for the non-linearity of the COSMOS magnitude scale 
proposed by Lumsden et al. (1997) and the extinction values of Burstein
\& Heiles (1984).
The intercluster sample has been obtained in the magnitude range $17-18.8$
and for consistency we limited also the other samples in the same magnitude 
range. 
\\
The upper panels of Figures \ref{fig:prevcomplex} and 
the left upper panel of \ref{fig:prevplates}   
show the histograms of the galaxies in the cluster and intercluster
samples, with superimposed the distribution expected for the corresponding
uniform sample. The lower panels of the same figures show the overdensity
$\displaystyle{{ N \over \bar{N}}  }$ as a function of the
comoving distance, where $N$ is the observed number of galaxies
and $\bar{N}$ is the number of objects expected in the volume occupied by the 
structure in the case of uniform distribution.
The plotted errors represent Poissonian 
uncertainties in the observed galaxy counts and a dotted line
corresponding to $\displaystyle{ { N \over \bar{N}}  = 1}$
has been drawn as a reference. 


The second step is the estimate of the volume occupied by the structure under
consideration. 
While the solid angle is assumed to be given by the surveyed area, the 
determination of the width of the supercluster in the direction
along the line of sight is more subjective. 
The data in the cluster complexes can not be used for this determination, 
because of the presence of significant peculiar velocities due to the 
virialization of clusters (also known as ``finger-of-God" effect), 
which broaden the width in the velocity space.
Therefore, we estimated the width of the supercluster by using only 
the distribution of galaxies in the intercluster survey, where the peculiar 
velocities are expected to be significantly smaller.  
\\
Since we find that the distribution of the distances from the best fit plane 
of the intercluster galaxies in the Shapley Concentration can be reasonably 
well described by a Gaussian function (see previous Section), we considered
as part of the supercluster all galaxies with distance from the plane smaller
than $\pm 2\sigma$.
Since the plane is tilted with respect to the ($\alpha$, $\delta$, $v$) 
reference frame, this choice produces a variable velocity range depending on
the position.
The minimum and maximum velocity ranges are $\Delta v \sim 2400$ km/s 
and $\Delta v \sim 3100$ km/s for fields $\# 48$ (plate 509) and $\# 5$ 
(plate 443), respectively.
\\
The physical width is assumed to be simply $\Delta v / H_o$, a choice 
acceptable in the case of low peculiar velocities. It is impossible to have an 
estimate of the peculiar
velocity pattern of this region, because of the difficulty of modelling 
the mass distribution: however, since it is reasonable to think that
this region (except the clusters) is not yet virialized, we expect that 
galaxies are still infalling toward the centre of the structure. 
We could parametrize this uncertainty as done by Small et
al. (1998a), which introduced the parameter $F$, defined as 
the ratio between the width of the supercluster in redshift space and in 
real space. In this case 
$\displaystyle{\left( N \over \bar{N} \right)_{real} = F 
\left( {N \over \bar{N} } \right)_{obs}  }$. 
\\
If the peculiar infall velocities (perpendicular to our fitted plane) were 
of the order of $\sim 150$ km/s,
similar to those measured in the Great Wall (Dell'Antonio, Geller \& Bothun 
1996), the value of $F$ for our structure would be $\sim 0.83$, so that 
our derived overdensities could be overestimated by $\sim 17\%$. 
Given the uncertainties on the amount of peculiar velocities, in the following 
we prefer to present our results neglecting the effect of $F$.
\\
Also in the cluster complexes and in A1736 we assumed that the width of the
supercluster is the same as that derived from our fit of the intercluster
galaxies, even if, due to the high peculiar motions induced by the virialized 
state of these clusters, these structures appear significantly elongated in
redshift space. 
In order to correct, at least partially, for this effect, we applied the
following procedure. First we assigned to the supercluster all objects with 
distance $\pm 2\sigma$ from the fitted plane; then we considered galaxies 
outside this velocity range: when they were in excess with respect to
the expected number, we assigned them to the supercluster. 
\\
We note that this procedure is somewhat arbitrary, in particular in presence 
of substructures infalling toward the clusters with high velocity. 
In the density profile plots these substructures may appear as secondary
peaks clearly separated  by the main overdensity. In these cases,
it is impossible to say if the observed difference in velocity between the
main and secondary peaks is due to a real spatial distance (and in this
case it would be not correct to assign galaxies of the substructure to the 
supercluster) or to a velocity difference. 
In our sample there are two clear cases of this kind. 
In the A3528 complex, there is the 
clump corresponding to A3535 at $v \sim 20000$ km/s, while in the A1736 cluster
we find the substructure already detected by Dressler \& Shectman (1988).
In order to derive a conservative estimate of the supercluster overdensity,
we choose to neglect these subclumps.  
\\
The definition of the limits of S300 is not very clear, 
because on plate 443 it consists of two features (see Figure \ref{fig:wedge}a), 
well separated by a void (labelled V3), while on plates 444 and 509 appears
as a single feature. 
In the plate 443 sample, the nearest peak (S300a) appears to extend 
from 28000 to 31000 km/s, while the velocity range of the farthest (S300b) 
overdensity is $[34000-38000]$ km/s. 
In the plate 444 and 509 samples, S300 appears as a single density excess
(S300c): 
we estimate a velocity range of $[30000-37000]$ km/s for the plate 444 sample 
and $[32000-35000]$ km/s for the plate 509 sample. 
\\
It is not clear if all these features are part of a single structure, but
on the other hand the distribution of galaxies in this region appears highly
coherent. For this reason, we computed a global overdensity in this region,
regardless of the physical association among the various components: therefore,
in the following we will present the results obtained considering the sum of
the contributions of S300a, S300b, V3 and S300c.   

\setcounter{table}{2} 
\begin{table*}
\caption[]{Estimated overdensities for the samples in the Shapley Concentration}
\begin{flushleft}
\begin{tabular}{lrrrrr}
\hline\noalign{\smallskip}
Sample & Overdensity  & Volume & Scale  & Mass & Redshift \\
 ~~ & $\displaystyle{ N \over \bar{N} }$ & [\htre] & [\hmpc]  & 
[$\Omega_o$ h$^{-1}$ M$_{\odot}$] & completeness  \\
\noalign{\smallskip}
\hline\noalign{\smallskip}
A3558 complex       & 46.4 $\pm$ 2.4 & ~~415.6 & ~4.6 & 5.4\ 10$^{15}$ & 64\% \\
A3528 complex       & 21.4 $\pm$ 1.2 & ~~609.5 & ~5.3 & 3.6\ 10$^{15}$ & 72\% \\
A1736               & 19.7 $\pm$ 3.1 & ~~247.8 & ~3.9 & 1.4\ 10$^{15}$ & 22\% \\
Intercluster sample & ~3.9 $\pm$ 0.4 & ~3082.8 & ~9.0 & 3.3\ 10$^{15}$ & 25\% \\
Total sample        & 11.3 $\pm$ 0.4 & ~4355.8 & 10.1 & 1.4\ 10$^{16}$ & 41\% \\
Extended sample     & ~6.8 $\pm$ 0.4 & 11611.3 & 14.1 & 2.2\ 10$^{16}$ & ~~~~ \\
Extended sample$+$382\&383     & ~5.2 $\pm$ 0.3 & 15607.8 & 15.5 & 2.3\ 10$^{16}$ & ~~~~ \\
\noalign{\smallskip}
\hline
\end{tabular}
\end{flushleft}
\label{tab:overdensity}
\end{table*}
\begin{table*}
\caption[]{Estimated overdensities in the S300 structure}
\begin{flushleft}
\begin{tabular}{llrrl}
\hline\noalign{\smallskip}
Sample & Overdensity  & Volume & Scale  & 
Mass \\
 ~~ & $\displaystyle{ N \over \bar{N} }$ & [\htre] & [\hmpc]  & 
[$\Omega_o$ h$^{-1}$ M$_{\odot}$] \\
\noalign{\smallskip}
\hline\noalign{\smallskip}
 S300a   & 4.5 $\pm$ 0.6 & ~7760.1 & 12.3 & 9.8\ 10$^{15}$ \\
 V3 void & 0.0           & 12506.4 & 14.4 & 0.0            \\
 S300b   & 3.8 $\pm$ 0.7 & 14806.7 & 15.2 & 1.6\ 10$^{16}$ \\
 S300c   & 3.2 $\pm$ 0.4 & 28783.2 & 19.0 & 2.6\ 10$^{16}$ \\
 Total   & 2.9 $\pm$ 0.3 & 63856.4 & 24.8 & 5.1\ 10$^{16}$ \\
\noalign{\smallskip}
\hline
\end{tabular}
\end{flushleft}
\label{tab:overd300}
\end{table*}

The third step is the computation of the overdensity $\displaystyle{ { N \over 
\bar{N}} }$ for each structure. 
The total overdensity in galaxies can then be estimated by combining the 
density excesses of the various samples following Postman, Geller \& Huchra 
(1988) as

\begin{equation}
\left( { N  \over \bar{N} } \right)_{SC} = \sum_i f_{i} 
\left( { N  \over \bar{N}} \right)_{i}\ , 
\label{eq:combine}
\end{equation}
where $f_i$ is the volume fraction occupied by the considered $i^{th}$
sample with overdensity $\displaystyle{\left( { N \over \bar{N}} \right)_{i}}$. 
\\
In Table \ref{tab:overdensity} the overdensities of the single 
structures (listed in column 1) and the total density excess in the Shapley 
Concentration are reported in column 2. 
The volumes and the redshift completeness of the samples are also 
reported (column 3 and 6).
In Table \ref{tab:overd300} we present the estimated overdensities and the 
volumes involved in the S300 samples.
 
In order to assign a formal scale to these overdensities, we calculated the 
radius of each structure assuming a spherical shape as

\begin{equation}
R=\left( {3 \over 4} {{V}\over{\pi}} \right)^{ {1\over 3}} \ ,
\end{equation}
where $R$ is the scale and $V$ is the volume.
In column 4 of Tables \ref{tab:overdensity} and \ref{tab:overd300}, 
the values of the scale for each structure in the various samples are reported.

\subsection{The mass estimate }

A direct measure of the mass in the supercluster regions between clusters
is not possible, because their small overdensities indicate that they are
far from virialization. For this reason we use the number excess in 
galaxies, which can be related to the mass excess (see below).

For what concerns the clusters, in previous studies from the literature their
contribution to the total mass or overdensity has been derived by adding 
together the results of the virial mass estimates of each cluster. 
In order to be fully consistent 
with the density excess derived for the intercluster survey, we chose 
to estimate also the cluster contribution as an excess in galaxy number. 
This procedure could be uncorrect if there is a significant segregation
between visible and dark matter in clusters with respect to the field, i.e. 
the proportionality between light and mass is different inside and outside
clusters.
However, the peculiar dynamical situation of the clusters 
in this region could also influence the reliability of the mass estimates.
Indeed, there is not agreement even for the mass of A3558, the richest and
best studied cluster of the Shapley Concentration. The mass
obtained from optical data ranges from $3.4 \times 10^{14}$  
h$^{-1}$ M$_{\odot}$ (Dantas et al. 1997) to $6 \times 10^{14}$  
h$^{-1}$ M$_{\odot}$ (Biviano et al. 1993), while the X-ray data
give masses in the range $1.5-6.0 \times 10^{14}$  
h$^{-1}$ M$_{\odot}$ (Bardelli et al. 1996, Ettori et al. 1997). 

We remind also that in the overdensity estimates we neglected the contribution
of a number of clusters (see Sect. 3.3), for which the available data are too
sparse for a density reconstruction.  
However, having rescaled the overdensity found for A1736 with the ratio
of the ACO richness parameters of these clusters, we found that neglecting
their contribution to the total overdensity of the Shapley Concentration leads 
to an underestimate of $\displaystyle{N \over { \bar{N}}}$ of the order of 
$10\%$.

Given the volume and the overdensity of a structure, the mass can be
estimated as

\begin{equation}
M= 2.778  \times 10^{11} \ \left({ { N} \over { {\bar N}}}\right)
 \ V  \  \Omega_o \ h^{-1} M_{\odot} \ , 
\label{eq:massa}
\end{equation}
where  $V$ is the volume (in \htre) occupied by the considered structure, 
$\displaystyle{{N}\over{{\bar N}}}$ is its number excess in galaxies with 
respect to a uniform distribution and $\Omega_o$ is the matter density 
parameter. This relation is correct 
only if the luminous matter traces exactly the distribution of the total matter.
The values found for the masses of the various structures by using 
eq.(\ref{eq:massa}) 
are reported in column 5 of Table \ref{tab:overdensity} and 
Table \ref{tab:overd300}.

As a more general case, we can assume (following Kaiser 1994) that the density 
excess in galaxies (or in clusters) is proportional to the total mass 
overdensity through the bias factor $b$

\begin{equation}
{{N - {\bar N}} \over {{\bar N}}} = 
b  \left( { { \rho - {\bar \rho}  } \over { {\bar \rho}}} \right) \ . 
\end{equation}
Hudson (1993a, 1993b), comparing the peculiar velocity field in the local 
Universe with the distribution of optical galaxies, estimated the most likely 
range for the bias factor to be approximately 
$2.5 \Omega_o^{0.6}-1.4 \Omega_o^{0.6}$. 
The first value comes from the comparison of the overdensity of the Virgo 
supercluster with the Virgo infall pattern, while the second value is obtained
by assuming that the large--scale galaxy dipole 
converges at a depth of $8000$ km/s. Therefore, the lower value could be 
an underestimate of $b$ if the dipole of the peculiar velocities has a deeper
depth (Plionis \& Valdarnini 1991; Scaramella et al. 1991). 
Indeed, an analysis of the dipole component perpendicular to the supergalactic
plane led to the higher value of $2.0 \Omega_o^{0.6}$ (Hudson 1993a).

%
%
\section{Results and discussion} 

The total overdensity of the Shapley Concentration in our surveyed region, 
computed following eq.(\ref{eq:combine}), is
$\displaystyle{ { N} \over { {\bar N}}}=11.3$ over a scale of $10.1$ \hmpc
(see Table \ref{tab:overdensity}).
In order to give an idea of the spatial pattern of the galaxy overdensity 
inside the Shapley Concentration, in the right panel of Figure 
\ref{fig:prevplates} we give the value of $\displaystyle{ {N}\over{{\bar N}}}$
in each MEFOS field: as expected, these values are higher in proximity of
the clusters. In the North-East part the overdensities decrease and there
are three fields with no observed galaxies in the considered velocity range:  
note however that since the expected number of galaxies in these fields is  
$\sim 0.8$, fields with no galaxies are still compatible with the 
mean density. Field \# 44 has the highest overdensity value 
(13.7), probably because of the presence of a group.  
\\
If we assume that the estimated overdensity of the intercluster galaxies 
extends also outside the MEFOS fields
over the region covered by plates 443, 444 and 509 (see the area delimited
by dashed lines in Figure \ref{fig:prevplates}), the total galaxy excess
derived from eq.(\ref{eq:combine}) is $6.8 \pm 0.4$ on a scale of $14.1$ \hmpc.
Hereafter we refer to this case as ``extended sample", while with the name
``total sample" we indicate the results for our original surveyed area. 
We remind that these samples include the contribution of the A3528 and A3558 
complexes and of A1736, but neglect the contribution of the other clusters
in the region (see Sect.3.3): therefore these overdensity values have to
be regarded as lower limits.  
\\
Drinkwater et al. (1999) give a value of $\displaystyle{ {N}\over{{\bar N}}}
= 2.0 \pm 0.2$ for plates 382 and 383, over an area of 44 sq.deg.; 
in order to add the contribution of this survey to our data, we computed its  
volume in the following way.
As shown in Figure \ref{fig:drinkwater}, our plane representation holds also 
in this sample; for the width determination we assume that this plane is 
perpendicular to the line of sight, therefore we adopt a width of 
$\pm 7.6$ \hmpc (see Sect.4.1). 
We added the contribution of these plates to our sample, following 
eq.(\ref{eq:combine}).
With these values we find for the Shapley Concentration a total overdensity 
of $\displaystyle{ {N}\over{{\bar N}}}=5.2 \pm 0.3$ on a scale of 15.5 \hmpc. 
Hereafter we refer to this case as ``extended sample $+$ 382 \& 383".
Also in this case, given the fact that the contribution of clusters in
plates 382 and 383 has been neglected, the overdensity values are lower limits.
\\
For what concerns the S300 structure, we have a total overdensity of
$\displaystyle{ { N} \over { {\bar N}}}=2.9$ over a scale of $24.8$ \hmpc:  
in Table \ref{tab:overd300} the various contributions to this structure
are reported.  However, we remark that it is not clear if all these features 
are physically part of a single structure: for this reason the values
of the reported overdensities have to be regarded as indicative of 
the matter distribution in this region.

Given these overdensities, 
the total mass of the Shapley Concentration in our surveyed region 
and of the S300 structure are  
$1.4\times 10^{16}$ $\Omega_o$ h$^{-1}$ M$_{\odot}$ and
$5.1\times 10^{16}$ $\Omega_o$ h$^{-1}$ M$_{\odot}$, respectively,
if light traces mass.  

The contribution of the Shapley Concentration to the peculiar velocity
of the Local Group with respect to the Cosmic Microwave Background reference
frame could be estimated as (Davis \& Peebles 1983)
\begin{equation}
\Delta v = 2.86\ 10^{-7} { { \Delta M  } \over { v^2 }} \Omega_o^{-0.4}\ h \ 
km/s,  
\end{equation}
where $\Delta M$ is the mass excess (in M$_\odot$) of the structure with 
respect to a uniform distribution and $v$ is the mean radial velocity
(in km/s). 
\\
Considering the ``extended $+$ 382 \& 383" sample, we find $\Delta v \sim 26$
km/s (for $\Omega_o=1$ and no bias): note however that this value corresponds 
only to the central part of the Shapley Concentration and does not take into
account the contribution of the matter in the external regions of the
supercluster.
\\
As a comparison, the peculiar velocity induced on the Local Group by the
Great Attractor is predicted to be of the order of $\sim 300$ km/s by
Hudson (1993a). 

Finally, it is possible to study how clusters and galaxies trace the same
structure. Assuming a mean cluster density of $25.2\times 10^{-6}$ clusters
Mpc$^{-3}$ (as found by Zucca et al. 1993 for the ACO clusters), we expect in 
the surveyed region of the Shapley Concentration (delimited by the dashed lines 
in Figure \ref{fig:prevplates}) 0.29 clusters, to be compared with the 11 
observed. This corresponds to an overdensity in clusters
of $\displaystyle{\left({ N \over \bar{N}}\right)_{cl}}=37.9\pm 11.4$. 
The quantity  
$b_{cl,g}= \displaystyle{
\left[\left({ N \over \bar{N}}\right)_{cl}-1 \right] / 
\left[\left({ N \over \bar{N}}\right)_{gal}-1 \right] 
}$ 
represents the ratio between the bias factors of clusters and galaxies. 
We found $ b_{cl,g}= 6.4 \pm 2.0$ on a scale of 14.1 \hmpc. 
Although the large associated error, this quantity seems to be inconsistent 
with the range of $2-3.5$ found for $b_{cl,g}$ comparing the dipoles (Branchini 
\& Plionis 1996) or the ratio of the correlation functions (see Scaramella 
et al. 1994) or the reconstructed linear power spectra (Peacock \& Dodds 1994)
of the distribution of clusters and galaxies.   
As reference, the same quantity for the Corona Borealis supercluster
is $b_{cl,g}=\displaystyle{7.53 \over 7}\sim 1$ on $\sim 20$ \hmpc, where the 
value for the galaxies is taken from Small, Sargent \& Hamilton (1998b).  
This fact, while is confirmimg the impression that the Shapley Concentration
has an anomalous richness in clusters with respect to other superclusters, 
on the other hand could raise problems for a simple comparison between results
on the large--scale distribution of clusters and galaxies, because
suggests significant variations of $b_{cl,g}$ with the local density.
However, in order to assess this result, detailed studies of the galaxy
distribution in a sample of superclusters are needed and are now possible with 
the new generation of multi--objects facilities.   

\subsection{Comparison with previous results}

The comparison of our values for Shapley Concentration and S300 with the 
results from the literature for
other superclusters is not straightforward, because of the different scales
over which the overdensities are computed. Moreover, sometimes the density 
excesses are given in {\it mass}, sometimes in {\it number} of galaxies or
clusters, therefore a direct comparison needs hypotheses about the value of
$\Omega_o$ and of the bias factor $b$. 
Moreover, these comparisons are affected by the relatively large uncertainties 
in the estimated values. 
However, it can be instructive to compare the results obtained using different
methods to estimate masses and overdensities, as f.i. virial masses, X--ray
masses, etc.
 
The overdensity for the Shapley Concentration (extended sample $+$ 382\&383)
can be compared with the density excess in {\it mass} given by Raychaudhury 
(1989) for the Great Attractor on a similar scale 
($\displaystyle {\rho \over{ {\bar \rho }}}=6 \Omega_o^{-1}$ over $15$ \hmpc).  
Under the hypothesis that $\Omega_o=1$ and that the galaxy distribution traces
that of the total matter, the two superclusters would be very similar on
the considered scale. 

\begin{figure}
\epsfysize=8.5cm
\epsfxsize=\hsize \epsfbox{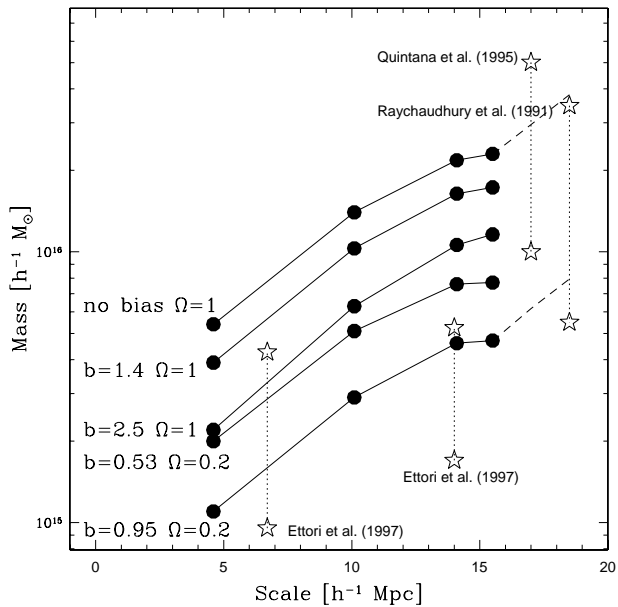} \hfil
\caption[]{ Estimated masses at various scales for different parameter choices,
indicated near each curve.
Stars connected with dotted lines refer to mass ranges given in the literature.
Dashed lines give the extrapolation of our mass values to the scales of
Quintana et al. and Raychaudhury et al., assuming the same overdensity we found
at our largest scale. }
\label{fig:masses}
\end{figure}

A more consistent comparison can be done with the list of superclusters
of Hudson (1993a), who studied the distribution of optical galaxies taken
from the UGC and ESO-Uppsala catalogues within $8000$ km/s. 
On scales comparable with those considered in our survey for the Shapley
Concentration, only the overdensities of the Virgo (Local)supercluster and 
the Fornax-Eridanus supercluster have been determined (see his table 10), 
finding $\displaystyle{ { N} \over { {\bar N}}}=2.70$ and $1.35$, respectively.
Comparisons with the other structures of Hudson (1993a), which are considered
at very different scales, are not straightforward: indeed, as can be noticed
from Tables \ref{tab:overdensity} and \ref{tab:overd300}, there are large 
variations of $\displaystyle{ { N} \over { {\bar N}}}$ inside superclusters 
when different scales are considered.

The value of $\displaystyle{ { N} \over { {\bar N}}}=3.1$ for the overdensity 
in {\it galaxies} in the Great Attractor on a scale of $\sim 30$ \hmpc 
(Dressler 1988) can be compared with the value of $\sim 2.9$ of S300, 
revealing the similarity of these two structures. Note,
however, that our value is a lower limit since we neglected the presence
of clusters in the S300 structure. 

\begin{table*}
\caption[]{Dynamical state of the structures}
\begin{flushleft}
\begin{tabular}{lrrrr}
\hline\noalign{\smallskip}
Sample & Linear Overdensity & Linear scale 
& Turnaround Time & Collapse Time \\
 ~~ & $\displaystyle{ N \over \bar{N} }$ & [\hmpc] & [$10^{9}$ h$^{-1}$ yrs] 
    &   [$10^{9}$ h$^{-1}$ yrs]  \\
\noalign{\smallskip}
\hline\noalign{\smallskip}
 A3558 complex                & 2.5 & 15.8 & ~3.8  & ~7.7 \\
 A3528 complex                & 2.4 & 13.9 & ~4.3  & ~8.5 \\
 Shapley Concentration        & 2.3 & 21.6 & ~4.9  & ~9.8 \\
 S300 structure               & 1.8 & 32.1 & 10.1  & 20.2 \\
\noalign{\smallskip}
\hline
\end{tabular}
\end{flushleft}
\label{tab:times}
\end{table*}

Marinoni et al. (1998), fitting the galaxy peculiar velocity field from the 
MarkIII catalogue (Willick et al. 1997) and the spiral galaxy sample of 
Matthewson, Ford \& Buchhorn (1992) with a multi--attractor model, 
found for the Shapley Concentration
$\displaystyle{ { \rho - {\bar \rho}  } \over { {\bar \rho}}} \sim$ 
2.7 and 3.8, respectively, within a scale of $15$ \hmpc: 
the comparison with our $\displaystyle{N \over \bar{N}}$ for the ``extended
$+$ 382\& 383" sample leads to values for the bias factor of
$b\sim$ 1.6 and 1.1, consistent with the adopted range of Hudson 
(1993a, 1993b). Note that Marinoni et al. (1998) used a model with spherical
symmetry, while the geometry of our sample is much more complicated:
however, if we use the Marinoni's values obtained for scales of 
$10$ \hmpc and $20$ \hmpc, the estimates for $b$ do not significantly change.  
These results mean that the value of the bias factor inside the Shapley
Concentration does not differ from those found in other regions of the
local Universe.

Adopting the Hudson's values for the bias parameter, the mass of the Shapley 
Concentration in our surveyed region results in the range $[6.3-10.3] \times 
10^{15}$ h$^{-1}$ M$_{\odot}$ in the case of $\Omega_o=1$ and  
$[2.9-5.1] \times 10^{15}$ h$^{-1}$ M$_{\odot}$ in the case of $\Omega_o=0.2$. 
The relative overdensity ranges are
$\displaystyle{ { \rho - {\bar \rho}  } \over { {\bar \rho}}} = [4.1-7.4]$ 
and $[10.8-19.4]$, respectively. The mass values for each scale and for
different choices of $b$ and $\Omega_o$ are shown in Figure \ref{fig:masses}.
\\ 
The mass of the Shapley Concentration can be compared with other estimates 
found in the literature, also shown in Figure \ref{fig:masses}. 
Raychaudhury et al. (1991), using the virial mass estimator, found 
$M_{vir} =3.5\times 10^{16}$ h$^{-1}$ M$_{\odot}$ on scales of $18.5$ \hmpc, 
and simply summing the cluster masses 
found $M =5.5\times 10^{15}$ h$^{-1}$ M$_{\odot}$ on the same
scale. This range is consistent with the more recent estimates of Quintana et 
al. (1995) on a scale of 17 \hmpc: also for these authors, the higher mass 
value is derived with the virial estimator and the lower one is the sum of 
cluster masses. 
These values can be compared with the mass we derived considering the 
``extended sample $+$ 382 \& 383" discussed above for the Shapley 
Concentration, that is $ 2.3 \times 10^{16}$ $\Omega_o$ h$^{-1}$ M$_{\odot}$
on a scale of 15.5 \hmpc (with no bias). 
Assuming that the overdensity derived for this
sample can be extended also to larger scales, we extrapolated the
expected mass values at the Quintana et al. and Raychaudhury et al. scales
in the two extreme cases for $b$ and $\Omega_o$: these extrapolations are shown 
as dashed lines in Figure \ref{fig:masses}. Our mass range is compatible with
both the Quintana et al. and the Raychaudhury et al. values. 
However note that our estimate at the
scale 15.5 \hmpc has to be regarded as a lower limit, because does not include 
the contribution of all clusters in the region. 
\\ 
On a scale of $\sim 14$ \hmpc Ettori et al. (1997), using various mass 
estimators, found values in the range $[1.7-5.2] \times 10^{15}$ h$^{-1}$ 
M$_{\odot}$. This range is consistent only with our lower estimate
for the ``extended sample" on scale 14.1 \hmpc: however, these authors took
into account only the clusters, neglecting the contribution of matter
outside clusters. 
\\ 
Ettori et al. (1997) gave a mass estimate also for the core of the supercluster,
ranging from $1.0 \times 10^{15}$ h$^{-1}$ M$_{\odot}$ to $4.3 \times 10^{15}$ 
h$^{-1}$ M$_{\odot}$ on a scale of $6.7$ \hmpc. On a comparable scale (4.6  
\hmpc on the A3558 complex), we find values in the range $[2.2-3.9] 
\times 10^{15}$ h$^{-1}$ M$_{\odot}$ in the case of bias and $\Omega_o=1$ and 
$[1.1-2.0] \times 10^{15}$ h$^{-1}$ M$_{\odot}$ in the case of bias and 
$\Omega_o=0.2$. 
The relative overdensity ranges are $\displaystyle { { \rho - {\bar \rho}  } 
\over { {\bar \rho}}} = [18.2-32.4]$ and $[47.8-85.7]$, respectively, to be
compared with the range $[4.05-11.47]$ given by Ettori et al. (1997). 
Also in this case, the discrepance between these ranges is probably due to 
the fact that these authors neglected the contribution of the matter outside 
clusters on the considered scale.

\subsection{Dynamics and comparison with theoretical models}

Rich superclusters are ideal laboratories where to study all the dynamical 
phenomena like cluster formation, because the high local densities induce
high peculiar velocities, which accelerate merging events. A standard method to
study the dynamical state of our structures is to follow the
evolution of their equivalent present linear overdensities (Kaiser \& Davis
1985; see Appendix A of Ettori et al. 1997 for the details of the formalism). 
In pratice, the linear overdensity and scale are the values that 
a structure would have if its gravitational evolution were linear.

In Table \ref{tab:times} we report the values of the linear density excess 
(column (2)), linear radius (column (3)) and turnaround and collapse times 
(columns (4) and (5)) in the case of $\Omega_o=1$ and no bias for the various 
structures (indicated in column (1)). The times have to be compared with the 
age of the Universe, which is $t_o=6.5\ 10^9$ h$^{-1}$ yrs (in the 
Einstein-deSitter case).  

From the values in Table \ref{tab:times} it is clear that, if light traces 
mass and $\Omega_o=1$, the Shapley Concentration already reached its turnaround 
radius and started to collapse: the final collapse will happen in $\sim 3
\ 10^9$ h$^{-1}$ yrs. We computed the turnaround times also in
the case of high and low bias ($b=2.5 \Omega_o^{0.6}$ and $b=1.4 \Omega_o^{0.6}$
respectively) and $\Omega_o=1$ and $\Omega_o=0.2$, finding that the Shapley
Concentration is still following a decelerated expansion in all cases, except
in the case of low bias and $\Omega_o=1$, for which the structure has just 
started to collapse.

For what concerns the S300 structure, it results it is well far from the
collapse in all considered scenarios.

The dynamical analysis of the cluster complexes indicates that the A3558 
complex is at the late stages of the collapse, which will happen in
$\sim 1\ 10^9$ h$^{-1}$ yrs. This result is consistent
with the analysis of the A3558 complex by Bardelli et al. (1994, 1998a, 1998b),
who found a complex dynamical situation, related to cluster mergings.
Analogous results about the A3528 complex will be presented in Bardelli et al.
(in preparation).

\begin{table*}
\caption[]{Comparison with theoretical models }
\begin{flushleft}
\begin{tabular}{lllllll}
\hline\noalign{\smallskip}
 & SCDM$_{COBE}$   & SCDM$_{CL}$  & $\tau$CDM  & TCDM  & OCDM & $\Lambda$CDM \\
\noalign{\smallskip}
\hline
\noalign{\smallskip}
 & \multicolumn{6}{l} {Shapley Concentration total sample}   \\
\noalign{\smallskip}
\hline
\noalign{\smallskip}
$\Omega_o=1$ no bias & {\it 3.07} & 7.22 & 5.79 & 6.49 &      &      \\
$\Omega_o=1$ $b=1.4$ & {\it 2.54} & 5.96 & 4.90 & 5.43 &      &      \\
$\Omega_o=1$ $b=2.5$ & {\it 1.81} & 4.26 & {\it 3.63} & {\it 3.93} &   &   \\
$\Omega_o=0.2$ no bias  &   &      &      &      & {\it 2.79} & {\it 2.58} \\
$\Omega_o=0.2$ $b=0.53$ &   &      &      &      & {\it 3.73} & {\it 3.43} \\
$\Omega_o=0.2$ $b=0.95$ &   &      &      &      & {\it 2.86} & {\it 2.63} \\
\noalign{\smallskip}
\hline
\end{tabular}
\end{flushleft}
\label{tab:models}
\end{table*}

Having calculated the linear scale and overdensity of the structures, it is
possible to compare them with the predictions of various theoretical models.
We considered a set of six different cosmological models belonging to the
general class of Cold Dark Matter (CDM) scenarios (see e.g. Moscardini et al. 
1998 ):
\\
1) the standard CDM model, with a normalization consistent with the COBE data
(SCDM$_{COBE}$);
\\
2) the same model but with a different normalization, in agreement with the
cluster abundances (SCDM$_{CL}$);
\\
3) the so--called $\tau$CDM model with a shape parameter $\Gamma = 0.21$;
\\
4) a tilted CDM model with $n=0.8$ (TCDM);
\\
5) an open CDM model with a matter density parameter $\Omega_o = 0.2$ (OCDM);
\\
6) a low density ($\Omega_o = 0.2$) CDM model with flatness given by the
cosmological constant ($\Lambda$CDM).
\\
All these models, except the first, are normalized by using the observed 
cluster abundance (see Eke, Cole \& Frenk 1996). Note that this normalization 
is also compatible with the COBE normalization, with the exception of
SCDM$_{CL}$, for which this normalization gives predicted values 
$\sim 43\%$ higher. For this reason, we present the results for SCDM with
both COBE and cluster normalizations. 

In Table \ref{tab:models} we report the ratio between the linear overdensity
and the r.m.s. mass fluctuations predicted by the various models for the
``total" sample and with different choices of $\Omega_o$ and $b$; the results
for the ``extended" and ``extended $+$ 382\&383" samples are similar. 
\\ 
In order to assess the statistical consistency of the Shapley Concentration
``event" with the predictions of the models, it is necessary to calculate
how many of such structures are expected in the sampled volume of the Universe.
Our survey is not useful to this purpose, because it covers a small solid angle 
just in the direction of the Shapley Concentration. Therefore, we decided to
consider the cluster sample used by Zucca et al. (1993) to search for 
superclusters: this sample covers the whole sky with $|b^{II}| > 15^o$ up to a
distance of $\sim 300$ \hmpc. 
In this volume there are no other superclusters as rich as the 
Shapley Concentration.
We computed the ratio between this volume and the volume of our samples listed
in Table \ref{tab:overdensity}: this corresponds to the number of available
``probes". Given this number, it is possible to estimate the number of $\sigma$
over which only 0.25 objects are expected in a Gaussian distribution. This
number of $\sigma$ is 4.33 for the ``total" sample. 
If the ratio between the linear 
overdensity and the r.m.s. mass fluctuations predicted by the various models 
exceeds this number of $\sigma$, we expect to have no event like the Shapley
Concentration in the sampled volume. 
\\ 
Therefore the existence of the Shapley Concentration is inconsistent with all 
models in Table \ref{tab:models} which predict values higher than $\sim 4$;
the consistent values are written in italic in Table \ref{tab:models}. 
The most discrepant model is the SCDM$_{CL}$; the TCDM and $\tau$CDM
give marginally consistent values only in case of high bias. On the contrary,
all models with a matter density parameter $<1$ seem to be compatible with the 
existence of 
a Shapley Concentration ``event". Finally, note that the SCDM$_{COBE}$   
appears to give the lowest values, but this model is inconsistent
with other results obtained on the scales of galaxies and clusters
(see e.g. Peacock \& Dodds 1996).
 
%
%
\section{Summary}

We have presented the results 
of a redshift survey of intercluster galaxies toward the central part
of the Shapley Concentration supercluster, aimed at determining the
distribution of galaxies in between obvious overdensities. Our sample is
formed by 442 new redshifts mainly in the $b_J$ magnitude range $17-18.8$.
Together with our redshift surveys on the A3558 and A3528 complexes,
our total sample has $\sim 2000$ velocities. 

Our main results are the following:
\\
-- The average velocity of the observed intercluster galaxies in the Shapley
Concentration appears to be a function of the ($\alpha$, $\delta$) position,  
and can be fitted by a plane in the three--dimensional space ($\alpha$, 
$\delta$, $v$): the distribution of the galaxy distances around the best fit 
plane is described by a Gaussian with a dispersion of $3.8$ \hmpc. 
\\ 
-- Using the 1440 galaxies of our sample in the magnitude range 
$17 - 18.8$, we reconstructed the density profile 
in the central part of the Shapley Concetration and we detected another 
significant overdensity at $\sim 30000$ km/s (dubbed S300). 
\\
-- We estimated the total overdensity in galaxies, the mass and the dynamical
state of these structures, discussing the effect of considering a bias 
between the galaxy distribution and the underlying matter. 
The estimated total overdensity in galaxies of these two structures 
is $\displaystyle{ N \over \bar{N}}\sim 11.3$ on scale of $10.1$ \hmpc
for the Shapley Concentration and  $\displaystyle{ N \over \bar{N}} \sim 2.9$
on scale of $24.8$ \hmpc for S300. If light traces the mass distribution, 
the corresponding masses are 
$1.4\times 10^{16}$ $\Omega_o$ h$^{-1}$ M$_{\odot}$ and 
$5.1\times 10^{16}$ $\Omega_o$ h$^{-1}$ M$_{\odot}$ for Shapley Concentration
and S300, respectively.
\\
The dynamical analysis revealed that, if light traces 
mass and $\Omega_o=1$, the Shapley Concentration already reached its turnaround 
radius and started to collapse: the final collapse will happen in $\sim 3
\ 10^9$ h$^{-1}$ yrs. 
\\
-- We compared our mass estimates on various scales with other results in 
the literature, finding a general agreement. 
\\
-- We found an indication that the value of the bias between clusters
and galaxies in the Shapley Concentration is higher than 
that reported in literature, confirming the impression that this supercluster
is very rich in clusters. 
\\
-- Finally from the comparison with some theoretical scenarios, we found that
the Shapley Concentration is more consistent with the predictions of the models
with a matter density parameter $<1$, such as open CDM and $\Lambda$CDM. 

%
%
\section*{Acknowledgements}

We warmly thank Andrea Biviano for having given us an electronic version of 
the redshift data on A1736 and Christian Marinoni for his overdensity data 
of the Shapley Concentration.

%
%

%

\setcounter{table}{1} 
\begin{table}
\caption[]{ Redshift data for the intercluster sample}
\begin{flushleft}
\begin{tabular}{rrrrrr}
\hline\noalign{\smallskip}
\multicolumn{6}{l}{ PLATE 443        } \\
\noalign{\smallskip}
\hline\noalign{\smallskip}
$\alpha$ (2000) & $\delta$ (2000) & $b_J$ & $v$ & err & notes \\
\noalign{\smallskip}
\hline\noalign{\smallskip}
  12~57~59.74 & -29~32~43.9 & 17.88 & 31721 & ~59 &          \\ 
  12~58~19.96 & -29~49~07.9 & 17.80 & 16879 & ~69 &          \\ 
  12~58~34.06 & -29~26~12.7 & 17.78 & 25656 & ~90 &          \\ 
  12~58~39.13 & -29~30~30.9 & 18.31 & 32048 & 171 &          \\ 
  12~58~53.31 & -29~19~49.8 & 18.66 & 32077 & ~50 &          \\ 
  12~59~22.10 & -29~09~37.2 & 18.68 & 37148 & 106 &          \\ 
  12~59~27.08 & -29~49~01.1 & 17.25 & 14779 & ~69 &          \\ 
  12~59~46.95 & -29~12~40.0 & 18.62 & 33036 & 145 &          \\ 
  13~00~18.07 & -29~44~04.1 & 17.56 & 17183 & ~46 &          \\ 
  13~00~35.04 & -28~53~36.2 & 18.60 & 20412 & ~14 & emiss    \\ 
  13~00~42.39 & -29~51~18.2 & 17.54 & 18381 & ~22 & emiss    \\ 
  13~00~46.28 & -29~43~35.0 & 17.45 & 32185 & 103 &          \\ 
  13~00~52.40 & -29~38~06.5 & 17.80 & 28528 & ~72 &          \\ 
  13~00~55.25 & -28~48~02.9 & 18.12 & 16301 & 146 &          \\ 
  13~01~02.65 & -28~15~58.9 & 18.56 & 40884 & ~61 &          \\ 
  13~01~07.29 & -29~54~22.6 & 18.40 & 16529 & ~78 &          \\ 
  13~01~08.13 & -28~59~58.2 & 18.17 & 19563 & ~38 &          \\ 
  13~01~14.71 & -29~09~07.5 & 17.99 & 25461 & ~84 &          \\ 
  13~01~21.41 & -28~34~23.2 & 18.35 & 22666 & 111 &          \\ 
  13~01~23.49 & -28~49~29.3 & 17.94 & 20113 & ~36 &          \\ 
  13~01~55.26 & -29~07~14.1 & 18.78 & 20128 & ~47 &          \\ 
  13~01~56.28 & -29~24~24.0 & 17.24 & 14821 & ~66 &          \\ 
  13~02~44.43 & -28~50~12.2 & 18.05 & 26028 & ~33 &          \\ 
  13~02~45.77 & -28~28~28.6 & 18.28 & 15304 & 146 &          \\ 
  13~02~48.45 & -28~45~15.7 & 18.64 & 23276 & 106 &          \\ 
  13~02~57.03 & -28~10~15.2 & 18.48 & 32293 & ~65 &          \\ 
  13~03~01.21 & -28~42~07.1 & 18.55 & 31959 & ~34 &          \\ 
  13~03~05.59 & -28~53~39.9 & 18.74 & 32412 & ~56 &          \\ 
  13~03~35.10 & -28~14~02.4 & 18.62 & 20459 & ~39 &          \\ 
  13~04~00.45 & -28~22~27.0 & 18.60 & 31361 & ~32 &          \\ 
  13~04~13.19 & -29~31~20.1 & 18.24 & 14562 & ~34 & emiss    \\ 
  13~04~27.74 & -29~33~54.2 & 17.46 & 15496 & ~10 & emiss    \\ 
  13~04~35.32 & -29~05~31.9 & 18.79 & 40663 & ~99 &          \\ 
  13~04~35.39 & -29~50~33.4 & 17.82 & 25772 & ~32 &          \\ 
  13~04~47.60 & -28~22~17.5 & 18.79 & 14685 & ~13 & emiss    \\ 
  13~04~49.98 & -30~44~48.0 & 17.52 & 15698 & ~35 &          \\ 
  13~04~52.26 & -28~36~13.6 & 18.61 & 14608 & ~10 & emiss    \\ 
  13~04~55.94 & -28~41~48.7 & 18.76 & 51978 & 107 &          \\ 
  13~05~04.65 & -29~11~12.1 & 18.03 & 31066 & ~40 &          \\ 
  13~05~08.80 & -28~23~45.1 & 17.13 & ~9427 & ~64 &          \\ 
  13~05~17.05 & -30~52~10.0 & 18.66 & 39399 & ~47 &          \\ 
  13~05~17.99 & -28~51~37.4 & 18.59 & 14998 & ~10 & emiss    \\ 
  13~05~26.26 & -28~58~13.9 & 18.43 & 31290 & ~46 &          \\ 
  13~05~27.80 & -29~10~10.4 & 17.53 & 30365 & ~81 &          \\ 
  13~05~31.58 & -29~54~00.2 & 18.07 & 20303 & ~82 &          \\ 
  13~05~31.93 & -29~24~59.5 & 17.81 & 19655 & ~23 &          \\ 
  13~05~32.83 & -28~24~39.2 & 18.09 & 22919 & ~60 &          \\ 
  13~05~35.15 & -31~02~31.0 & 17.58 & 32296 & ~81 &          \\ 
  13~05~41.99 & -30~57~11.3 & 17.50 & 29902 & ~59 &          \\ 
\noalign{\smallskip}
\hline
\end{tabular}
\end{flushleft}
\label{tab:sample}
\end{table}

\setcounter{table}{1} 
\begin{table}
\caption[]{ cont. }
\begin{flushleft}
\begin{tabular}{rrrrrr}
\hline\noalign{\smallskip}
\multicolumn{6}{l}{ PLATE 443        } \\
\noalign{\smallskip}
\hline\noalign{\smallskip}
  $\alpha$ (2000) & $\delta$ (2000) & $b_J$ & $v$ & err & notes \\
\noalign{\smallskip}
\hline\noalign{\smallskip}
 13~05~45.85 & -29~08~40.0 & 17.56 & 31561 & ~31 &          \\ 
 13~05~47.53 & -29~20~08.5 & 18.39 & 40349 & ~64 &          \\ 
 13~05~50.25 & -30~36~09.9 & 18.73 & 31889 & ~66 &          \\ 
 13~05~50.52 & -30~32~28.4 & 18.71 & 39476 & 119 &          \\ 
 13~05~53.44 & -28~38~56.0 & 18.66 & 31314 & ~71 &          \\ 
 13~05~53.98 & -28~19~59.2 & 17.38 & ~2266 & ~35 & emiss    \\ 
 13~05~54.06 & -29~34~06.5 & 18.72 & 22248 & ~37 &          \\ 
 13~06~02.67 & -29~11~47.7 & 17.06 & 15063 & ~42 &          \\ 
 13~06~20.00 & -31~05~42.7 & 17.24 & ~3783 & ~97 &          \\ 
 13~06~24.54 & -28~42~56.1 & 18.54 & 48985 & 101 &          \\ 
 13~06~24.57 & -29~33~02.1 & 18.60 & 20478 & ~64 &          \\ 
 13~06~24.82 & -28~14~12.8 & 17.93 & 38966 & ~44 &          \\ 
 13~06~24.82 & -30~14~22.6 & 17.17 & 15490 & 105 &          \\ 
 13~06~25.95 & -29~26~49.9 & 18.76 & 40258 & ~30 &          \\ 
 13~06~27.09 & -28~22~29.6 & 18.37 & 32462 & ~52 &          \\ 
 13~06~37.92 & -29~38~22.1 & 17.18 & 19754 & ~45 &          \\ 
 13~06~47.48 & -30~11~43.7 & 17.74 & ~3443 & 107 &          \\ 
 13~06~52.04 & -28~56~04.8 & 17.48 & 21435 & ~75 &          \\ 
 13~06~55.60 & -28~39~49.3 & 17.38 & 31893 & ~89 &          \\ 
 13~06~55.87 & -28~53~13.7 & 18.67 & 26429 & ~30 & emiss    \\ 
 13~07~00.19 & -28~18~24.8 & 18.61 & ~6330 & ~60 & emiss    \\ 
 13~07~04.05 & -29~08~00.6 & 18.64 & 35006 & 122 &          \\ 
 13~07~06.70 & -28~49~17.9 & 18.57 & 34453 & ~33 &          \\ 
 13~07~13.90 & -30~35~15.6 & 17.14 & ~3320 & ~10 & emiss    \\ 
 13~07~16.76 & -28~30~49.1 & 17.33 & 22266 & 262 &          \\ 
 13~07~24.04 & -31~07~27.1 & 18.41 & 36270 & ~46 &          \\ 
 13~07~25.90 & -30~22~21.7 & 18.11 & 22439 & ~88 &          \\ 
 13~07~35.84 & -28~48~01.1 & 18.52 & 31765 & ~37 & emiss    \\ 
 13~07~37.01 & -30~14~23.0 & 17.17 & 16919 & 107 &          \\ 
 13~07~41.51 & -30~52~58.6 & 18.66 & 39162 & ~42 &          \\ 
 13~07~51.58 & -30~39~18.2 & 18.34 & 14210 & ~45 &          \\ 
 13~07~53.05 & -29~16~40.4 & 18.21 & 31960 & ~87 &          \\ 
 13~07~55.93 & -29~36~56.2 & 18.78 & 35135 & ~27 &          \\ 
 13~07~56.66 & -28~18~43.1 & 18.65 & 40655 & 186 &          \\ 
 13~07~58.35 & -28~29~15.5 & 18.11 & 31823 & ~30 &          \\ 
 13~08~10.83 & -28~34~25.5 & 17.81 & 32473 & ~65 &          \\ 
 13~08~12.01 & -30~33~52.2 & 18.58 & 16928 & ~10 & emiss    \\ 
 13~08~20.29 & -29~58~48.1 & 18.73 & 52649 & ~42 &          \\ 
 13~08~22.15 & -29~23~40.7 & 18.70 & 30982 & ~98 &          \\ 
 13~08~29.51 & -30~22~43.2 & 18.67 & 49202 & 135 &          \\ 
 13~08~40.32 & -28~35~56.6 & 18.44 & 32380 & ~54 &          \\ 
 13~08~49.52 & -29~19~04.3 & 18.16 & 32544 & ~66 &          \\ 
 13~09~02.48 & -30~46~34.7 & 17.21 & 11779 & ~82 &          \\ 
 13~09~06.31 & -29~43~06.8 & 18.11 & 49031 & ~50 &          \\ 
 13~09~07.85 & -30~37~47.2 & 18.11 & ~9540 & ~10 & emiss    \\ 
 13~09~15.44 & -28~42~47.9 & 18.14 & 32389 & ~51 &          \\ 
 13~09~19.59 & -29~32~55.1 & 17.60 & ~2886 & ~39 &          \\ 
 13~09~25.16 & -29~18~01.3 & 17.44 & 14861 & ~97 &          \\ 
 13~09~33.38 & -29~43~21.3 & 17.34 & 14948 & ~10 & emiss    \\ 
\noalign{\smallskip}
\hline
\end{tabular}
\end{flushleft}
\end{table}

\setcounter{table}{1} 
\begin{table}
\caption[]{ cont. }
\begin{flushleft}
\begin{tabular}{rrrrrr}
\hline\noalign{\smallskip}
\multicolumn{6}{l}{ PLATE 443        } \\
\noalign{\smallskip}
\hline\noalign{\smallskip}
  $\alpha$ (2000) & $\delta$ (2000) & $b_J$ & $v$ & err & notes \\
\noalign{\smallskip}
\hline\noalign{\smallskip}
 13~09~45.60 & -29~12~42.2 & 18.00 & 23714 & ~46 &          \\ 
 13~09~47.44 & -30~52~22.3 & 18.36 & 16908 & ~21 &          \\ 
 13~09~48.01 & -28~22~06.8 & 17.78 & 14304 & 102 &          \\ 
 13~09~49.11 & -29~20~34.1 & 18.62 & 39497 & ~43 &          \\ 
 13~09~50.06 & -29~31~00.1 & 18.05 & 14869 & ~36 &          \\ 
 13~09~54.92 & -29~24~43.5 & 18.63 & 39219 & ~54 &          \\ 
 13~09~54.95 & -30~00~59.2 & 17.09 & 16639 & ~77 &          \\ 
 13~09~56.38 & -30~40~11.3 & 17.11 & 13409 & 117 &          \\ 
 13~10~00.18 & -30~26~54.0 & 18.69 & 31188 & ~53 &          \\ 
 13~10~06.42 & -28~27~56.6 & 18.02 & 48899 & ~52 &          \\ 
 13~10~11.02 & -28~19~47.5 & 18.39 & 35627 & ~73 &          \\ 
 13~10~19.83 & -28~38~46.5 & 18.59 & 39393 & ~95 &          \\ 
 13~10~20.51 & -29~01~15.9 & 18.62 & 39458 & 185 & emiss    \\ 
 13~10~30.45 & -30~38~07.4 & 18.16 & ~9782 & ~28 &          \\ 
 13~10~34.46 & -30~26~55.2 & 18.77 & 48245 & ~47 &          \\ 
 13~10~37.37 & -29~01~12.0 & 18.38 & 32543 & ~62 &          \\ 
 13~10~42.20 & -28~50~19.9 & 18.18 & ~4650 & ~10 & emiss    \\ 
 13~10~42.53 & -29~06~35.3 & 18.29 & 30853 & ~45 &          \\ 
 13~10~42.65 & -30~50~27.3 & 18.30 & 35574 & ~78 &          \\ 
 13~10~44.32 & -29~59~33.9 & 18.74 & 38941 & ~31 &          \\ 
 13~10~50.36 & -28~42~25.2 & 18.62 & 36365 & ~50 &          \\ 
 13~10~53.75 & -28~36~57.0 & 18.49 & 40191 & ~37 &          \\ 
 13~10~54.69 & -30~23~31.1 & 18.62 & 22539 & 117 &          \\ 
 13~10~56.85 & -29~38~32.4 & 18.63 & ~3229 & ~16 &          \\ 
 13~10~59.46 & -29~58~16.7 & 18.76 & 38504 & ~61 &          \\ 
 13~11~02.03 & -30~57~45.7 & 18.02 & 19827 & ~10 &          \\ 
 13~11~07.14 & -29~16~11.2 & 18.52 & 49460 & ~56 & emiss    \\ 
 13~11~19.30 & -29~49~37.8 & 18.57 & 52514 & ~79 &          \\ 
 13~11~22.03 & -30~20~54.4 & 17.88 & 39861 & ~72 &          \\ 
 13~11~28.25 & -28~56~18.8 & 18.58 & 49488 & ~75 &          \\ 
 13~11~29.59 & -28~47~05.8 & 18.52 & 31992 & ~39 &          \\ 
 13~11~35.25 & -29~48~14.6 & 17.78 & 30573 & ~79 &          \\ 
 13~11~45.55 & -29~36~38.2 & 18.30 & ~3529 & ~88 &          \\ 
 13~11~53.55 & -30~16~04.1 & 18.19 & 24274 & ~47 &          \\ 
 13~12~08.62 & -28~30~04.4 & 17.08 & ~6969 & ~51 & emiss    \\ 
 13~12~18.11 & -30~45~55.9 & 17.02 & 35933 & ~35 &          \\ 
 13~12~33.26 & -30~47~31.2 & 17.75 & 35947 & ~66 &          \\ 
 13~12~40.15 & -28~38~18.9 & 18.25 & 18321 & 138 &          \\ 
 13~12~47.16 & -28~25~22.9 & 18.23 & 23431 & ~10 & emiss    \\ 
 13~12~49.85 & -31~21~31.8 & 17.49 & 20899 & ~76 &          \\ 
 13~12~59.19 & -30~30~58.0 & 18.21 & 32540 & ~62 &          \\ 
 13~13~02.70 & -30~25~53.9 & 18.69 & ~9776 & ~41 &          \\ 
 13~13~07.82 & -31~34~37.5 & 18.75 & 32331 & ~38 &          \\ 
 13~13~09.96 & -30~41~47.9 & 17.83 & 23302 & ~53 &          \\ 
 13~13~15.62 & -31~01~06.1 & 18.52 & 28809 & 130 &          \\ 
 13~13~17.28 & -28~30~23.2 & 18.66 & 38398 & ~82 &          \\ 
 13~13~30.74 & -30~37~18.1 & 18.53 & 35966 & ~57 &          \\ 
 13~13~35.12 & -31~48~20.8 & 17.68 & 14211 & ~10 & emiss    \\ 
 13~13~49.61 & -31~57~17.6 & 18.12 & 15585 & ~36 &          \\ 
\noalign{\smallskip}
\hline
\end{tabular}
\end{flushleft}
\end{table}

\setcounter{table}{1} 
\begin{table}
\caption[]{ cont. }
\begin{flushleft}
\begin{tabular}{rrrrrr}
\hline\noalign{\smallskip}
\multicolumn{6}{l}{ PLATE 443        } \\
\noalign{\smallskip}
\hline\noalign{\smallskip}
  $\alpha$ (2000) & $\delta$ (2000) & $b_J$ & $v$ & err & notes \\
\noalign{\smallskip}
\hline\noalign{\smallskip}
 13~13~56.53 & -32~06~11.9 & 18.02 & 34670 & ~73 &          \\ 
 13~13~57.70 & -30~49~05.7 & 18.56 & 47538 & ~65 &          \\ 
 13~14~09.50 & -31~35~37.5 & 17.69 & 31555 & ~42 &          \\ 
 13~15~26.60 & -31~16~41.7 & 17.82 & 31812 & ~25 &          \\ 
 13~15~33.91 & -31~20~13.5 & 18.42 & 31195 & 159 &          \\ 
 13~15~40.58 & -31~29~20.1 & 18.69 & 32782 & ~39 &          \\ 
 13~15~42.42 & -31~25~22.9 & 18.44 & 38262 & ~31 & emiss    \\ 
 13~15~55.03 & -31~42~51.0 & 17.50 & ~4612 & ~18 & emiss    \\ 
 13~16~04.26 & -31~51~11.2 & 18.76 & 38378 & ~39 &          \\ 
 13~16~19.67 & -31~35~04.6 & 18.07 & 39184 & ~36 &          \\ 
\noalign{\smallskip}
\hline
\end{tabular}
\end{flushleft}
\end{table}

\setcounter{table}{1} 
\begin{table}
\caption[]{ cont. }
\begin{flushleft}
\begin{tabular}{rrrrrr}
\hline\noalign{\smallskip}
\multicolumn{6}{l}{ PLATE 444  } \\
\noalign{\smallskip}
\hline\noalign{\smallskip}
   $\alpha$ (2000) & $\delta$ (2000) & $b_J$ & $v$ & err & notes \\
\noalign{\smallskip}
\hline\noalign{\smallskip}
 13~14~23.51 &    -30~39~34.7 &    18.78 &    11998 & ~10 &  emiss\\
 13~14~24.78 &    -28~42~32.9 &    18.21 &    23693 & 153 &       \\
 13~14~40.55 &    -30~41~55.9 &    18.76 &    36147 & 101 &       \\
 13~14~43.03 &    -30~50~21.4 &    17.49 &    14728 & 101 &       \\
 13~14~47.94 &    -30~24~11.3 &    18.78 &    30509 & 109 &       \\
 13~14~51.06 &    -28~52~54.4 &    18.62 &    39765 & ~51 &       \\
 13~14~58.55 &    -30~16~21.4 &    18.19 &    21479 & ~40 &       \\
 13~15~07.98 &    -30~25~27.6 &    18.30 &    38954 & ~34 &       \\
 13~15~08.14 &    -30~57~02.6 &    18.27 &    29750 & ~53 &       \\
 13~15~18.05 &    -30~33~51.4 &    18.04 &    38888 & ~41 &       \\
 13~15~32.58 &    -30~20~19.2 &    18.21 &    25858 & ~50 &       \\
 13~15~51.53 &    -28~44~43.8 &    18.04 &    30333 & 177 &       \\
 13~15~56.21 &    -28~29~05.0 &    17.22 &    13825 & ~73 &       \\
 13~15~59.98 &    -30~10~38.9 &    17.78 &    16101 & ~10 &  emiss\\
 13~16~01.64 &    -30~31~15.2 &    18.26 &    72825 & ~46 &       \\
 13~16~13.41 &    -30~48~18.1 &    18.11 &    25746 & ~61 &       \\
 13~16~17.09 &    -30~43~19.6 &    18.55 &    36348 & ~74 &       \\
 13~16~37.62 &    -30~55~26.2 &    17.75 &    30339 & ~72 &       \\
 13~16~41.01 &    -30~34~56.4 &    17.76 &    15297 & ~42 &  emiss\\
 13~17~04.19 &    -30~06~15.5 &    18.14 &    30360 & ~67 &       \\
 13~17~14.57 &    -30~44~26.5 &    17.35 &    21536 & ~25 &       \\
 13~17~22.58 &    -30~23~58.9 &    18.10 &    29991 & ~50 &       \\
 13~17~24.19 &    -28~30~37.1 &    18.10 &    15750 & ~72 &       \\
 13~18~00.56 &    -28~36~15.6 &    17.69 &    37396 & ~39 &       \\
 13~18~36.24 &    -28~27~41.9 &    18.04 &    37569 & ~90 &       \\
 13~19~24.70 &    -30~53~14.2 &    18.11 &    50426 & ~23 &  emiss\\
 13~19~30.12 &    -30~26~13.6 &    17.65 &    27762 & ~37 &       \\
 13~19~34.29 &    -30~29~10.2 &    18.40 &    29264 & ~98 &       \\
 13~19~40.72 &    -30~57~24.5 &    18.24 &    17968 & ~40 &       \\
 13~19~49.52 &    -30~21~02.4 &    18.32 &    34194 & ~79 &       \\
 13~20~31.94 &    -30~34~54.7 &    17.19 &    23387 & ~71 &       \\
 13~20~33.45 &    -30~32~13.1 &    18.48 &    35016 & ~29 &       \\
 13~20~39.99 &    -30~45~43.9 &    17.56 &    33082 & ~42 &       \\
 13~20~44.59 &    -30~20~43.7 &    18.00 &    ~4456 & ~73 &       \\
 13~20~47.66 &    -30~28~28.7 &    18.32 &    51391 & ~37 &       \\
 13~20~53.50 &    -30~12~24.3 &    17.85 &    39614 & ~50 &       \\
 13~21~08.11 &    -30~34~47.6 &    18.48 &    33156 & ~37 &       \\
 13~21~12.67 &    -29~53~27.2 &    17.29 &    14556 & ~35 &       \\
 13~21~24.11 &    -30~23~45.3 &    18.24 &    35484 & ~48 &       \\
 13~21~31.10 &    -30~45~54.6 &    18.32 &    49642 & ~72 &       \\
 13~21~35.10 &    -30~51~36.6 &    18.46 &    14492 & ~27 &  emiss\\
 13~21~36.18 &    -30~10~49.9 &    18.25 &    13854 & ~54 &       \\
 13~21~43.25 &    -29~39~05.8 &    17.71 &    23289 & ~81 &       \\
 13~21~53.22 &    -30~02~24.0 &    18.43 &    17154 & ~20 &  emiss\\
 13~21~55.42 &    -30~39~03.2 &    18.31 &    33141 & ~35 &       \\
 13~22~01.61 &    -29~56~54.0 &    17.84 &    34945 & ~37 &       \\
 13~22~02.15 &    -30~06~18.1 &    17.36 &    14800 & ~76 &       \\
 13~22~03.77 &    -29~35~11.8 &    17.00 &    ~9877 & ~32 &       \\
 13~22~17.92 &    -30~09~15.7 &    17.57 &     2489 & ~10 &  emiss\\
 13~22~22.28 &    -30~13~20.1 &    18.10 &    33261 & ~58 &       \\
\noalign{\smallskip}
\hline
\end{tabular}
\end{flushleft}
\end{table}

\setcounter{table}{1} 
\begin{table}
\caption[]{ cont. }
\begin{flushleft}
\begin{tabular}{rrrrrr}
\hline\noalign{\smallskip}
\multicolumn{6}{l}{ PLATE 444         } \\
\noalign{\smallskip}
\hline\noalign{\smallskip}
   $\alpha$ (2000) & $\delta$ (2000) & $b_J$ & $v$ & err & notes \\
\noalign{\smallskip}
\hline\noalign{\smallskip}
 13~22~33.65 &    -29~18~26.6 &    18.17 &    23032 & ~90 &       \\
 13~22~37.69 &    -30~25~12.6 &    17.77 &    25741 & ~32 &       \\
 13~22~44.86 &    -29~38~10.1 &    18.22 &    38377 & ~83 &       \\
 13~22~49.66 &    -30~06~51.8 &    18.14 &    ~4286 & ~13 &  emiss\\
 13~23~00.49 &    -29~18~32.9 &    17.77 &    14389 & 110 &       \\
 13~23~20.83 &    -29~42~26.2 &    17.77 &    25798 & ~61 &       \\
 13~23~40.68 &    -30~00~51.8 &    18.62 &    ~9396 & ~16 &  emiss\\
 13~23~41.04 &    -29~33~24.4 &    18.68 &    22351 & ~15 &       \\
 13~23~49.77 &    -29~55~05.4 &    18.47 &    45851 & 132 &       \\
 13~23~51.58 &    -29~21~41.2 &    18.50 &    22990 & ~47 &       \\
 13~24~24.97 &    -29~42~10.8 &    17.35 &    ~9845 & 158 &       \\
 13~24~39.07 &    -30~40~50.7 &    17.91 &    14533 & 100 &       \\
 13~24~48.35 &    -30~32~10.3 &    17.73 &    27469 & ~54 &       \\
 13~24~52.66 &    -30~36~28.4 &    18.53 &    27234 & ~34 &       \\
 13~24~55.63 &    -29~52~58.1 &    18.60 &    32789 & ~90 &       \\
 13~24~55.65 &    -30~44~51.4 &    17.57 &    27310 & ~35 &       \\
 13~24~55.76 &    -31~02~26.3 &    17.79 &    15658 & 101 &       \\
 13~25~07.02 &    -29~59~50.1 &    17.75 &    14005 & ~35 &       \\
 13~25~07.08 &    -29~30~23.2 &    18.03 &    13925 & ~26 &       \\
 13~25~12.84 &    -30~20~21.4 &    17.77 &    ~4374 & ~20 &  emiss\\
 13~25~23.87 &    -29~49~58.9 &    18.09 &    14490 & ~83 &       \\
 13~25~23.95 &    -30~54~06.7 &    17.20 &    14171 & 144 &       \\
 13~25~29.56 &    -30~45~34.9 &    17.70 &    43708 & ~41 &       \\
 13~25~35.52 &    -30~20~10.1 &    17.53 &    14050 & ~13 &  emiss\\
 13~26~34.06 &    -30~11~43.7 &    18.27 &    44029 & ~76 &       \\
 13~26~34.51 &    -30~47~32.6 &    18.51 &    14464 & ~30 &  emiss\\
 13~26~37.04 &    -31~00~22.7 &    17.95 &    15692 & ~10 &  emiss\\
 13~26~47.62 &    -30~28~59.9 &    17.34 &    ~9957 & ~12 &  emiss\\
 13~26~49.84 &    -30~37~56.6 &    17.90 &    14555 & ~68 &       \\
 13~27~00.37 &    -30~14~40.6 &    17.31 &    ~4224 & ~10 &  emiss\\
 13~27~29.06 &    -30~40~15.7 &    17.12 &    14136 & ~31 &       \\
 13~27~29.55 &    -30~33~31.4 &    17.13 &    14150 & ~10 &  emiss\\
 13~27~34.32 &    -30~23~03.3 &    18.26 &    13901 & ~16 &  emiss\\
 13~27~36.15 &    -30~58~47.7 &    17.92 &    13093 & ~79 &       \\
 13~28~03.33 &    -30~40~50.1 &    17.81 &    14996 & ~55 &       \\
 13~28~06.31 &    -30~34~49.8 &    17.80 &    14362 & ~43 &       \\
 13~28~23.03 &    -28~18~14.7 &    17.20 &    ~2053 & ~20 &  emiss\\
 13~28~26.44 &    -28~47~21.2 &    18.75 &    13366 & ~31 &  emiss\\
 13~28~28.63 &    -30~38~27.1 &    18.16 &    37888 & ~30 &       \\
 13~28~30.26 &    -30~46~55.2 &    18.33 &    37905 & ~36 &       \\
 13~28~41.59 &    -28~23~55.2 &    17.44 &    10792 & ~96 &       \\
 13~28~44.52 &    -30~34~09.1 &    18.76 &    38859 & ~93 &       \\
 13~28~46.53 &    -30~57~08.2 &    18.62 &    36656 & ~72 &       \\
 13~28~47.63 &    -30~21~23.9 &    18.20 &    27081 & ~36 &       \\
 13~28~58.94 &    -28~34~42.5 &    18.31 &    10722 & ~95 &       \\
 13~29~15.94 &    -28~26~06.9 &    18.13 &    ~9228 & ~20 &  emiss\\
 13~29~20.35 &    -30~46~37.7 &    17.97 &    14398 & ~94 &       \\
 13~29~31.15 &    -28~10~26.7 &    18.17 &    ~9793 & ~24 &       \\
 13~29~35.47 &    -28~57~09.6 &    18.22 &    42651 & ~49 &       \\
\noalign{\smallskip}
\hline
\end{tabular}
\end{flushleft}
\end{table}

\setcounter{table}{1} 
\begin{table}
\caption[]{ cont. }
\begin{flushleft}
\begin{tabular}{rrrrrr}
\hline\noalign{\smallskip}
\multicolumn{6}{l}{ PLATE 444        } \\
\noalign{\smallskip}
\hline\noalign{\smallskip}
   $\alpha$ (2000) & $\delta$ (2000) & $b_J$ & $v$ & err & notes \\
\noalign{\smallskip}
\hline\noalign{\smallskip}
 13~29~36.44 &    -30~32~15.7 &    18.08 &    14562 & ~33 &       \\
 13~29~44.29 &    -29~00~19.0 &    17.67 &    14694 & ~25 &       \\
 13~29~46.59 &    -30~48~06.3 &    18.26 &    43569 & ~47 &       \\
 13~29~56.15 &    -30~29~33.8 &    18.45 &    27398 & 133 &       \\
 13~29~57.30 &    -28~14~27.1 &    18.15 &    ~9627 & ~12 &  emiss\\
 13~30~03.86 &    -31~05~41.5 &    17.75 &    13726 & ~28 &       \\
 13~30~05.47 &    -30~55~34.7 &    18.06 &    43402 & ~37 &       \\
 13~30~08.06 &    -28~37~17.5 &    18.76 &    34699 & ~82 &  emiss\\
 13~30~15.84 &    -30~36~17.4 &    17.03 &    15319 & ~29 &       \\
 13~30~19.55 &    -28~05~04.3 &    17.45 &    10205 & ~39 &       \\
 13~30~31.31 &    -30~19~29.7 &    18.01 &    37319 & ~61 &       \\
 13~30~31.84 &    -28~08~24.1 &    18.36 &    14195 & ~27 &  emiss\\
 13~30~42.61 &    -31~04~07.5 &    17.21 &    16085 & ~93 &       \\
 13~30~49.33 &    -30~19~23.6 &    18.36 &    40717 & ~34 &       \\
 13~30~53.17 &    -28~14~05.7 &    18.53 &    ~9055 & ~86 &       \\
 13~30~55.99 &    -29~00~51.8 &    18.15 &    ~2173 & ~10 &  emiss\\
 13~30~56.30 &    -28~04~56.2 &    17.72 &    ~5650 & 189 &       \\
 13~31~05.59 &    -30~56~34.2 &    17.54 &    15781 & ~34 &       \\
 13~31~08.71 &    -30~40~44.8 &    17.77 &    26782 & ~35 &       \\
 13~31~15.84 &    -28~52~52.6 &    17.62 &    13940 & 150 &       \\
 13~31~21.04 &    -30~45~54.9 &    17.64 &    22825 & ~31 &       \\
 13~31~21.93 &    -30~28~49.0 &    17.92 &    14506 & ~28 &       \\
 13~31~32.49 &    -30~55~13.1 &    17.53 &    14183 & ~34 &       \\
 13~31~37.96 &    -30~31~02.7 &    17.75 &    15003 & 102 &       \\
 13~31~44.35 &    -28~28~38.9 &    18.12 &    44286 & ~83 &       \\
 13~31~59.37 &    -30~42~57.9 &    18.08 &    15176 & 102 &       \\
 13~32~22.19 &    -28~37~14.7 &    18.43 &    13791 & ~32 &  emiss\\
 13~32~52.32 &    -28~37~52.4 &    17.24 &    34811 & ~29 &       \\
 13~33~09.99 &    -30~20~25.6 &    17.90 &    24342 & ~23 &  emiss\\
 13~33~13.22 &    -30~37~38.2 &    17.54 &    33162 & ~66 &       \\
 13~33~32.51 &    -30~12~00.2 &    17.10 &    14832 & ~76 &       \\
 13~33~35.80 &    -30~20~14.4 &    17.03 &    14698 & ~38 &       \\
 13~33~42.42 &    -28~29~43.0 &    17.70 &    13868 & ~20 &  emiss\\
 13~33~42.52 &    -30~28~09.9 &    17.14 &    10877 & ~20 &       \\
 13~33~42.97 &    -30~08~58.0 &    18.54 &    47984 & ~94 &       \\
 13~33~48.65 &    -28~15~25.1 &    18.60 &    ~9721 & ~46 &       \\
 13~33~55.53 &    -30~42~28.7 &    18.34 &    22400 & ~77 &       \\
 13~34~01.58 &    -30~49~26.5 &    18.26 &    37269 & ~67 &       \\
 13~34~10.74 &    -30~57~27.0 &    18.09 &    13635 & ~10 &  emiss\\
 13~34~19.45 &    -28~52~34.1 &    17.25 &    22656 & ~37 &       \\
 13~34~24.56 &    -28~41~11.1 &    18.33 &    34534 & ~59 &       \\
 13~34~38.57 &    -28~35~36.0 &    18.49 &    35852 & ~29 &  emiss\\
 13~34~40.86 &    -30~13~00.4 &    17.37 &    10995 & ~58 &       \\
 13~34~42.90 &    -30~26~02.7 &    17.46 &    15132 & ~82 &       \\
 13~35~04.11 &    -28~58~23.5 &    18.71 &    13262 & ~81 &       \\
 13~35~09.30 &    -30~11~59.1 &    18.25 &    35140 & 100 &       \\
 13~35~19.88 &    -30~18~58.5 &    17.42 &    22488 & ~10 &  emiss\\
 13~35~22.59 &    -30~49~22.7 &    18.35 &    39636 & ~53 &       \\
 13~35~46.54 &    -29~00~32.7 &    18.24 &    45965 & ~28 &       \\
 13~35~52.58 &    -30~47~27.2 &    18.20 &    22514 & ~58 &       \\
\noalign{\smallskip}
\hline
\end{tabular}
\end{flushleft}
\end{table}

\setcounter{table}{1} 
\begin{table}
\caption[]{ cont. }
\begin{flushleft}
\begin{tabular}{rrrrrr}
\hline\noalign{\smallskip}
\multicolumn{6}{l}{ PLATE 444        } \\
\noalign{\smallskip}
\hline\noalign{\smallskip}
   $\alpha$ (2000) & $\delta$ (2000) & $b_J$ & $v$ & err & notes \\
\noalign{\smallskip}
\hline\noalign{\smallskip}
 13~35~53.92 &    -28~17~58.4 &    17.18 &    13513 & ~11 &  emiss\\
 13~35~55.36 &    -30~33~11.1 &    17.29 &    ~4187 & ~27 &  emiss\\
 13~35~59.05 &    -30~22~34.4 &    18.28 &    13840 & ~10 &  emiss\\
 13~36~05.99 &    -28~26~08.7 &    18.37 &    35100 & ~29 &       \\
 13~36~19.57 &    -28~56~08.3 &    17.75 &    14439 & 114 &       \\
 13~36~23.50 &    -28~45~01.2 &    18.70 &    46416 & 118 &       \\
 13~36~29.19 &    -28~27~47.6 &    18.21 &    35020 & ~41 &       \\
 13~36~30.31 &    -28~34~48.5 &    17.77 &    35723 & ~36 &       \\
 13~36~37.84 &    -30~30~55.8 &    17.40 &    22269 & 111 &       \\
 13~36~41.41 &    -30~25~18.6 &    17.00 &    11319 & ~45 &       \\
 13~36~55.64 &    -28~40~31.3 &    18.66 &    46502 & 139 &       \\
\noalign{\smallskip}
\hline
\end{tabular}
\end{flushleft}
\end{table}

\setcounter{table}{1} 
\begin{table}
\caption[]{ cont. }
\begin{flushleft}
\begin{tabular}{rrrrrr}
\hline\noalign{\smallskip}
\multicolumn{6}{l}{ PLATE 509  } \\
\noalign{\smallskip}
\hline\noalign{\smallskip}
  $\alpha$ (2000) & $\delta$ (2000) & $b_J$ & $v$ & err & notes \\
\noalign{\smallskip}
\hline\noalign{\smallskip}
 13~25~10.03 &    -24~30~24.9 &    18.43 &    32843 & ~40 &       \\
 13~25~31.76 &    -24~25~03.7 &    17.54 &    22802 & ~66 &       \\
 13~25~31.96 &    -24~18~56.9 &    17.74 &    24183 & ~74 &       \\
 13~25~36.96 &    -25~46~45.1 &    17.13 &    ~9538 & 181 &       \\
 13~25~40.35 &    -25~57~45.1 &    18.75 &    19718 & ~14 &  emiss\\
 13~25~42.43 &    -25~53~12.4 &    17.41 &    18019 & ~36 &       \\
 13~25~43.19 &    -25~36~06.5 &    18.65 &    27159 & 187 &       \\
 13~25~45.61 &    -24~44~10.7 &    17.48 &    10086 & ~77 &       \\
 13~25~49.54 &    -24~34~13.0 &    17.43 &    12123 & ~30 &  emiss\\
 13~25~52.99 &    -24~12~50.0 &    18.55 &    12336 & ~18 &  emiss\\
 13~26~01.68 &    -26~08~12.2 &    18.34 &    26637 & ~43 &       \\
 13~26~17.94 &    -25~37~22.8 &    18.18 &    12514 & ~10 &  emiss\\
 13~26~20.15 &    -24~13~18.3 &    18.40 &    41508 & ~94 &       \\
 13~26~22.52 &    -25~56~25.7 &    18.16 &    31868 & ~42 &       \\
 13~26~25.60 &    -25~24~45.1 &    17.63 &    ~7716 & ~42 &       \\
 13~26~43.87 &    -25~18~42.4 &    18.33 &    50105 & ~82 &       \\
 13~26~44.62 &    -24~17~48.2 &    17.75 &    ~9821 & ~10 &  emiss\\
 13~26~45.49 &    -25~44~56.4 &    18.31 &    15436 & ~31 &       \\
 13~26~47.44 &    -26~10~41.1 &    18.31 &    39962 & ~43 &       \\
 13~26~47.98 &    -24~39~42.1 &    18.02 &    13483 & ~32 &       \\
 13~26~54.01 &    -24~46~27.1 &    18.36 &    13782 & ~58 &       \\
 13~26~56.43 &    -23~18~08.0 &    17.62 &    13633 & ~42 &       \\
 13~26~56.68 &    -24~16~11.9 &    17.66 &    13337 & ~36 &       \\
 13~27~04.23 &    -24~30~18.6 &    17.46 &    13502 & ~54 &       \\
 13~27~06.47 &    -25~27~46.6 &    17.51 &    13712 & ~44 &       \\
 13~27~09.10 &    -24~26~07.3 &    18.51 &    13723 & ~32 &       \\
 13~27~15.08 &    -24~01~47.0 &    17.65 &    34321 & ~49 &       \\
 13~27~32.39 &    -25~35~13.3 &    18.59 &    ~9075 & ~15 &  emiss\\
 13~27~36.02 &    -23~34~06.5 &    17.49 &    ~9745 & ~41 &       \\
 13~27~40.01 &    -24~12~42.7 &    17.19 &    ~9940 & ~25 &       \\
 13~27~43.55 &    -24~45~05.1 &    18.45 &    32693 & 134 &       \\
 13~27~46.98 &    -25~18~16.4 &    17.50 &    12955 & ~28 &       \\
 13~27~50.62 &    -24~56~06.4 &    17.58 &    27239 & ~38 &       \\
 13~27~51.87 &    -24~03~37.4 &    18.25 &    ~9338 & ~95 &       \\
 13~27~57.53 &    -25~40~14.3 &    17.95 &    11392 & ~76 &       \\
 13~27~58.62 &    -26~09~59.1 &    18.76 &    32308 & 133 &       \\
 13~28~01.56 &    -24~14~32.3 &    18.02 &    13296 & ~26 &       \\
 13~28~02.38 &    -26~02~37.5 &    18.26 &    36880 & 148 &       \\
 13~28~14.09 &    -23~39~27.1 &    18.34 &    41532 & ~70 &       \\
 13~28~14.56 &    -24~37~00.3 &    17.16 &    13360 & ~34 &       \\
 13~28~16.78 &    -24~21~10.7 &    17.93 &    12749 & ~30 &       \\
 13~28~18.22 &    -24~42~34.6 &    17.40 &    14149 & ~52 &       \\
 13~28~19.32 &    -25~57~42.8 &    18.53 &    30767 & 185 &       \\
 13~28~21.34 &    -25~28~35.7 &    17.27 &    13635 & ~31 &       \\
 13~28~22.50 &    -24~28~19.9 &    17.25 &    13486 & ~34 &       \\
 13~28~32.49 &    -24~00~30.4 &    17.17 &    12758 & ~61 &       \\
 13~28~38.58 &    -26~04~27.8 &    18.02 &    13473 & ~41 &       \\
 13~28~45.81 &    -24~10~22.9 &    18.71 &    12700 & ~42 &       \\
 13~28~46.29 &    -25~37~00.4 &    17.70 &    12649 & ~10 &  emiss\\
 13~29~12.54 &    -25~32~34.0 &    17.57 &    13060 & ~94 &       \\
\noalign{\smallskip}
\hline
\end{tabular}
\end{flushleft}
\end{table}

\setcounter{table}{1} 
\begin{table}
\caption[]{ cont. }
\begin{flushleft}
\begin{tabular}{rrrrrr}
\hline\noalign{\smallskip}
\multicolumn{6}{l}{ PLATE 509        } \\
\noalign{\smallskip}
\hline\noalign{\smallskip}
   $\alpha$ (2000) & $\delta$ (2000) & $b_J$ & $v$ & err & notes \\
\noalign{\smallskip}
\hline\noalign{\smallskip}
 13~29~25.19 &    -25~52~00.7 &    17.94 &    27692 & ~33 &       \\
 13~29~25.46 &    -23~39~34.8 &    18.17 &    ~5422 & ~10 &  emiss\\
 13~29~26.77 &    -23~44~23.6 &    18.74 &    23057 & ~48 &       \\
 13~30~07.83 &    -25~09~59.0 &    17.43 &    10448 & ~50 &       \\
 13~30~13.67 &    -25~05~33.2 &    18.19 &    32264 & ~81 &       \\
 13~30~29.81 &    -25~17~51.1 &    18.23 &    38231 & ~57 &       \\
 13~30~42.02 &    -25~21~25.1 &    17.28 &    13445 & ~94 &       \\
 13~30~50.03 &    -25~26~05.4 &    17.61 &    13556 & ~42 &       \\
 13~31~02.04 &    -24~57~39.0 &    18.62 &    32226 & ~91 &       \\
 13~31~21.56 &    -25~23~36.9 &    17.73 &    27941 & ~34 &       \\
 13~31~48.40 &    -24~56~01.2 &    17.43 &    ~7860 & ~63 &       \\
 13~32~04.35 &    -25~00~03.5 &    18.37 &    38543 & 250 &       \\
 13~32~35.44 &    -24~49~28.9 &    17.64 &    14880 & 123 &       \\
 13~32~43.60 &    -25~16~34.5 &    18.40 &    34941 & ~64 &       \\
 13~32~45.92 &    -25~24~11.8 &    18.45 &    35690 & ~70 &       \\
 13~32~55.69 &    -25~09~49.8 &    17.14 &    12429 & ~72 &       \\
 13~32~58.67 &    -25~23~29.0 &    18.68 &    71679 & ~71 &       \\
 13~33~43.10 &    -25~02~42.3 &    17.74 &    19161 & ~31 &       \\
 13~34~45.65 &    -24~41~02.6 &    18.61 &    36344 & ~46 &       \\
 13~34~45.99 &    -24~45~18.7 &    18.65 &    19208 & ~13 &  emiss\\
 13~34~48.52 &    -26~45~37.4 &    18.46 &    10914 & 111 &       \\
 13~34~55.38 &    -27~07~15.0 &    17.67 &    10533 & ~67 &       \\
 13~34~56.82 &    -27~00~59.1 &    17.44 &    ~9746 & ~58 &  emiss\\
 13~34~57.99 &    -24~22~28.8 &    18.29 &    38124 & ~64 &       \\
 13~35~05.00 &    -24~56~13.2 &    18.63 &    37110 & ~31 &       \\
 13~35~05.91 &    -27~14~32.7 &    18.15 &    10282 & ~42 &  emiss\\
 13~35~20.57 &    -25~47~55.3 &    18.37 &    46970 & 183 &       \\
 13~35~36.18 &    -25~40~01.0 &    17.02 &    31094 & 201 &       \\
 13~35~43.17 &    -27~11~13.6 &    18.27 &    11319 & ~18 &  emiss\\
 13~35~51.92 &    -25~57~38.2 &    18.44 &    28385 & ~49 &  emiss\\
 13~35~52.58 &    -24~41~08.4 &    17.56 &    20177 & ~74 &       \\
 13~35~55.62 &    -24~19~02.6 &    17.00 &    19596 & ~25 &  emiss\\
 13~35~59.00 &    -24~07~22.9 &    17.90 &    37849 & ~82 &       \\
 13~36~08.10 &    -25~27~23.3 &    17.89 &    32532 & ~45 &       \\
 13~36~21.74 &    -26~55~41.1 &    17.93 &    36770 & ~63 &       \\
 13~36~22.53 &    -26~46~45.3 &    17.78 &    35672 & ~62 &       \\
 13~36~24.27 &    -26~06~30.0 &    18.21 &    20469 & ~88 &       \\
 13~36~26.89 &    -27~14~07.7 &    17.26 &    13448 & ~97 &       \\
 13~36~29.28 &    -25~24~13.8 &    18.30 &    39324 & 151 &       \\
 13~36~36.63 &    -26~32~29.5 &    17.77 &    ~6865 & ~33 &       \\
 13~36~37.89 &    -27~03~17.7 &    17.74 &    13443 & ~85 &       \\
 13~36~40.65 &    -26~57~00.6 &    18.52 &    36272 & ~67 &       \\
 13~36~42.14 &    -24~22~47.9 &    18.46 &    37480 & ~69 &       \\
 13~36~54.89 &    -24~40~18.6 &    17.63 &    ~4762 & ~42 &       \\
 13~37~07.34 &    -24~10~04.0 &    18.32 &    36516 & ~43 &       \\
 13~37~13.48 &    -26~58~26.1 &    18.61 &    22094 & ~53 &       \\
 13~37~21.91 &    -25~48~14.2 &    18.62 &    25195 & 160 &       \\
 13~37~33.22 &    -27~15~19.9 &    18.00 &    33022 & ~36 &       \\
 13~37~37.41 &    -25~19~11.7 &    18.10 &    35883 & 135 &       \\
 13~37~38.86 &    -24~40~54.9 &    18.48 &    ~6752 & ~10 &  emiss\\
\noalign{\smallskip}
\hline
\end{tabular}
\end{flushleft}
\end{table}

\setcounter{table}{1} 
\begin{table}
\caption[]{ cont. }
\begin{flushleft}
\begin{tabular}{rrrrrr}
\hline\noalign{\smallskip}
\multicolumn{6}{l}{ PLATE 509        } \\
\noalign{\smallskip}
\hline\noalign{\smallskip}
   $\alpha$ (2000) & $\delta$ (2000) & $b_J$ & $v$ & err & notes \\
\noalign{\smallskip}
\hline\noalign{\smallskip}
 13~37~41.61 &    -24~29~16.6 &    18.16 &    20013 & ~53 &       \\
 13~37~46.97 &    -24~15~04.0 &    18.24 &    32058 & ~26 &       \\
 13~37~55.87 &    -26~41~57.8 &    17.79 &    32287 & 121 &       \\
 13~38~00.39 &    -25~53~16.4 &    18.25 &    30775 & ~73 &       \\
 13~38~08.45 &    -24~29~49.4 &    18.68 &    42571 & ~30 &       \\
 13~38~21.49 &    -25~23~11.4 &    17.77 &    35326 & ~99 &       \\
 13~38~22.69 &    -25~37~00.8 &    18.13 &    37322 & ~69 &       \\
 13~38~37.96 &    -27~07~12.1 &    17.15 &    ~4725 & ~24 &  emiss\\
 13~38~39.73 &    -26~45~53.7 &    18.06 &    26857 & ~71 &       \\
 13~41~55.79 &    -24~31~52.5 &    18.66 &    36573 & ~68 &       \\
 13~41~56.94 &    -24~23~01.0 &    17.41 &    16897 & 123 &       \\
 13~42~13.25 &    -24~09~25.5 &    18.02 &    37697 & ~97 &       \\
 13~42~16.89 &    -24~36~15.8 &    17.85 &    29093 & ~64 &       \\
 13~42~31.15 &    -24~47~26.7 &    17.59 &    18330 & 101 &       \\
 13~42~32.70 &    -24~31~06.0 &    18.71 &    36609 & ~73 &       \\
 13~42~47.60 &    -24~27~50.5 &    18.51 &    19372 & ~12 &  emiss\\
 13~43~25.14 &    -25~01~31.1 &    17.46 &    14530 & 114 &       \\
 13~43~32.11 &    -24~33~54.0 &    17.00 &    36728 & ~34 &       \\
 13~43~39.07 &    -24~20~12.2 &    17.40 &    ~6913 & ~50 &       \\
 13~43~53.74 &    -24~53~19.6 &    18.11 &    14341 & ~70 &       \\
 13~44~20.97 &    -24~50~53.4 &    17.70 &    43024 & ~91 &       \\
 13~44~22.65 &    -24~45~29.2 &    18.57 &    37839 & 165 &       \\
 13~44~37.25 &    -24~20~14.2 &    18.55 &    ~6369 & ~92 &       \\
 13~44~44.42 &    -24~42~58.3 &    18.60 &    36252 & 165 &       \\
 13~44~55.45 &    -24~37~30.0 &    18.17 &    27656 & ~28 &       \\
\noalign{\smallskip}
\hline
\end{tabular}
\end{flushleft}
\end{table}

\end{document}